\documentclass[12pt,preprint]{aastex}
\def\kms{\ifmmode{\rm km\thinspace s^{-1}}\else km\thinspace s$^{-1}$\fi}
\def\ms{\ifmmode{\rm m\thinspace s^{-1}}\else m\thinspace s$^{-1}$\fi}

\usepackage{amsmath,amsfonts}

\begin{document}

\title{High-resolution spectroscopic follow-up of OGLE planetary
transit candidates in the Galactic bulge: two possible Jupiter-mass
planets and two blends}

\author{Maciej Konacki\altaffilmark{1}}
\affil{Department of Geological and Planetary Sciences, California
Institute of Technology, MS 150-21, Pasadena, CA 91125, USA\\
Nicolaus Copernicus Astronomical Center, Polish Academy of Sciences, 
Rabia\'nska 8, 87-100 Toru\'n, Poland}

\author{Guillermo Torres\altaffilmark{2}, Dimitar D. Sasselov\altaffilmark{3}}
\affil{Harvard-Smithsonian Center for Astrophysics, 60 Garden St.,
Cambridge, MA 02138, USA}

\author{Saurabh Jha\altaffilmark{4}}
\affil{Department of Astronomy, University of California, Berkeley, CA
94720, USA}
\affil{Harvard-Smithsonian Center for Astrophysics, 60 Garden St.,
Cambridge, MA 02138, USA}

\altaffiltext{1}{e-mail: maciej@gps.caltech.edu}
\altaffiltext{2}{e-mail: gtorres@cfa.harvard.edu}
\altaffiltext{3}{e-mail: dsasselov@cfa.harvard.edu}
\altaffiltext{4}{e-mail: saurabh@astron.berkeley.edu}

\begin{abstract} 

We report the results of our campaign to follow-up spectroscopically
several candidate extrasolar transiting planets from the OGLE-III
survey in the direction of the Galactic center, announced in 2001
\citep{Udalski:02a::,Udalski:02b::}. All of these objects present
shallow and periodic dips in brightness that may be due to planetary
companions.  Our high-resolution Keck~I/HIRES observations have
revealed two interesting cases (OGLE-TR-10 with a period of 3.1~days,
and OGLE-TR-58 with a period of 4.3~days) that show no radial velocity
variations at the level of 100--200 $\ms$. If orbited by companions,
their masses would be similar to Jupiter. With the information in hand
(including the light curves) we are not able to rule out that these
candidates are instead the result of contamination from an eclipsing
binary in the same line of sight (a ``blend"). We discuss also the
case of OGLE-TR-56 that was recently reported by \cite{Konacki:03::}
to have a Jupiter-size companion, based on an earlier analysis of our
data, and we present supporting information.  Two other candidates,
OGLE-TR-3 and OGLE-TR-33, show clear evidence that they are blends.
We describe tests carried out to characterize the
stability of the HIRES spectrograph and its impact on the
determination of precise velocities for faint stars ($V\geq 15$~mag)
using exposures of
a Thorium-Argon lamp as the wavelength reference.  Systematic effects
are at the level of 100 $\ms$ or smaller, and tend to dominate the
total error budget. We also evaluate the precision attainable using
the iodine gas absorption cell as an alternative fiducial, and we
propose a simplified version of the standard procedure employed for
high-precision Doppler planet searches that is very promising. Results
from both this method and the classical Th-Ar technique show the
feasibility of spectroscopic follow-up for faint targets in the range
$V = 14$--17. 
	
\end{abstract}


\section{Introduction}

The application of high-precision radial velocity and pulsar timing
techniques over the past decade has resulted in the discovery of more
than 100 extrasolar giant planet candidates around solar-type stars
\citep{Schneider:03::}, and also smaller Earth-mass planets around a
millisecond pulsar \citep{Konacki:03b::,Wolszczan:92::}. With the
photometric confirmation in 1999 of the giant planet around HD~209458
through the detection of its transits
\citep{Henry:00::,Charbonneau:00::} it has become clear that this is a
viable search technique, applicable even to Earth-size planets in
future space missions. Numerous programs using both small wide-field
telescopes for bright stars and large-aperture telescopes for fainter
stars have been established to monitor increasing numbers of objects
with high photometric precision ($< 1$\%) in a variety of stellar
populations \citep[see, e.g.,][and references therein]{Horne:03::}.
The Optical Gravitational Lensing Experiment in its most recent
incarnation (OGLE-III) is among the most successful of these searches,
having uncovered 59 candidate transiting planets in its first
observing season (2001) in three fields toward the Galactic center
\citep{Udalski:02a::,Udalski:02b::}, another 62 in three other fields
in the constellation of Carina during 2002 \citep{Udalski:02c::}, and
16 additional candidates in the bulge and in Carina from a recent
reanalysis of the original photometric observations
\citep{Udalski:03::}.  All of these candidates show dips in the
brightness of the star of only a few percent, and have photometric
periods of a few days. Follow-up of the bulge sample has recently
resulted in the spectroscopic confirmation of a second extrasolar
transiting planet, around the star OGLE-TR-56, the first case
originally discovered from its photometric signature rather than its
Doppler signature \citep{Konacki:03::}. 

The success with HD~209458 ($V = 7.65$) has led to general optimism
about the possibility of detecting significant numbers of transiting
planets among similarly bright stars, and performing additional
observations of the kind carried out for HD~209458
\citep[e.g.,][]{Brown:01::}, which among other things led to the
remarkable detections of the atmosphere of the planet
\citep{Charbonneau:02::,Vidal:03::}. More skeptical views have been
voiced about the feasibility of high-precision Doppler follow-up of
faint targets such as those uncovered by the OGLE project and others
($V = 15$--19).  It has also been recognized that a variety of other
phenomena can mimic transit light curves by producing shallow
eclipses, and that these situations need to be examined carefully in
order to be ruled out before a planet discovery can be claimed.
Examples of these ``false positives" include a stellar (as opposed to
a substellar) companion orbiting a large star (B-A main sequence star,
or a giant), grazing eclipses in a stellar binary, and contamination
by the light of a fainter eclipsing binary along the same line of
sight (referred to as a ``blend"). 

So far no other transiting planets have been discovered among bright
stars. OGLE-TR-56 is a much fainter object ($V = 16.6$).  Furthermore,
from the results by \cite{Konacki:03::} there are indications that the
incidence of blends in searches among fainter stars in generally
crowded fields is rather high (possibly as high as 98\%).  Thus,
discovering additional transiting extrasolar planets has turned out to
be somewhat more difficult than anticipated \citep[see
also][]{Charbonneau:03::}. 

In this paper we report our findings for several more targets from the
OGLE-III sample towards the bulge, based on precise measurements of
the radial velocities of these stars. We present the methodology and
tests that document the stability of our instrumental setup, a key
issue for confirming the reality of any Doppler signal.  Results are
given for OGLE-TR-10 and OGLE-TR-58, which remain good candidates for
having a planetary companion in orbit around them, and additional
supporting details for OGLE-TR-56 are given to supplement those
reported by \cite{Konacki:03::}.  We discuss also the case of
OGLE-TR-3, recently claimed by \cite{Dreizler:03::} to harbor a
transiting planet, but we conclude that the evidence is actually
against that interpretation, and more in favor of a blend. Another
case of a blend, OGLE-TR-33, will be presented in full detail
elsewhere. 

\section{Observations}

Because the OGLE targets are relatively faint, high-resolution
follow-up of large numbers of them to detect their Doppler signatures
becomes expensive or impractical in terms of telescope time. Thus, our
strategy was to make use of all the information available from the
photometry to eliminate obvious binary systems, and then to perform
preliminary low-resolution spectroscopic observations
(``reconnaissance") in order to rule out others, so that we may focus
our attention for the high-precision velocity measurements only on the
best candidates. 
	
\subsection{Inspection of the light curves and low-resolution
spectroscopic reconnaissance}

Light curves based on the photometry from the OGLE-III survey for all
59 targets in the direction of the Galactic bulge were examined
carefully for any indications that might rule out the object as a good
candidate.  For example, we looked for signs of a secondary eclipse
and/or out-of-eclipse variations that are strongly indicative of a
massive (stellar) companion as opposed to a substellar companion. 

Our initial inspection of the photometry allowed us to rule out a
total of 20 candidates for a variety of reasons. Eight of them showed
fairly obvious signs of stellar secondaries in the light curves.  Four
have only one transit detected, and thus the period is unknown.
Another star is a duplicate entry (OGLE-TR-29 = OGLE-TR-8). A further
7 were considered too faint for follow-up and were also excluded, thus
leaving us with 39 candidates from the original list. It should be
noted here that since the time of our original inspection of the light
curves, more sophisticated numerical methods have been developed to
quantify their shape outside of eclipse \citep{Drake:03::,Sirko:03::}.
These lead to the rejection of a few more of the bulge candidates,
although these same objects were also rejected on the basis of the
low-resolution spectroscopy we describe in the following. 

The 39 best candidates were observed spectroscopically at relatively
low resolution with two different instruments. For the brighter
objects we used the 1.5-m Tillinghast reflector at the F.\ L.\ Whipple
Observatory on Mt.\ Hopkins (Arizona) in May and June 2002, with the
FAST spectrograph (resolving power $R \simeq 4400$). For the
remaining objects we used the 6.5-m Baade telescope (Magellan~I) at
the Las Campanas Observatory (Chile) on July 17--21 2002 with the
Boller \& Chivens spectrograph ($R \simeq 2200$).  One of the goals
was to identify the early-type stars among the candidates, which are
too large for an orbiting Jovian-size planet to produce a detectable
drop in brightness, implying that the companion must be stellar. A
handful of these cases were found in similar spectroscopic work at
somewhat lower resolution by \cite{Dreizler:02::} in the same sample.
A further goal was to reject stellar binaries with grazing eclipses by
measuring the radial velocities of the target stars multiple times.
Stellar binaries will show variations of tens of $\kms$ due to orbital
motion (which is large compared to the 2--3 $\kms$ precision we
achieved in these observations), as opposed to changes of only a few
hundred $\ms$ for true planets.  As a result of this reconnaissance,
we established that 8 of the candidates are of early spectral type
(B--A), and 25 showed velocity variations or even double lines in the
spectra indicating that they are stellar binaries. Only 6 showed no
significant variations within the errors, and were assigned the
highest priority for the high-resolution follow-up described below.
Further details of the low-resolution observations and complete
information about all the rejected OGLE candidates will be reported
elsewhere (Torres et al., in preparation). 

\subsection{High-resolution spectroscopy and radial velocity
measurements}

Five of our 6 best candidates were observed spectroscopically on 4
nights in July 24--27 2002 with the Keck~I telescope and the HIRES
instrument \citep{Vogt:94::}. These stars are OGLE-TR-3, OGLE-TR-10,
OGLE-TR-33, OGLE-TR-56 and OGLE-TR-58. The setup allowed us to record
35 usable echelle orders covering the spectral range from 3850~\AA\ to
6200~\AA\ at a resolving power of $R \simeq 65,\!000$. Typical
signal-to-noise ratios are in the range of 10--20 per pixel for a
single exposure.  Our main wavelength reference was provided by a
hollow-cathode Thorium-Argon lamp, of which we obtained short
exposures immediately preceding and following each stellar exposure.
For two of the brighter objects (OGLE-TR-10 and OGLE-TR-58) we also
made use of the iodine ($I_2$) gas absorption cell ---the standard
setup for highly-precise Doppler planet searches with this instrument
\citep[e.g.,][]{Vogt:00::}--- which imprints a dense spectrum of $I_2$
lines directly on the stellar spectrum in the region from
approximately 5000~\AA\ to 6200~\AA\ (some 12 echelle orders), and
which serves as a very accurate and stable reference frame for
measuring velocity shifts. Template exposures of these two stars
without the iodine were also taken on one of the nights for use in the
reduction process, although in the final iodine reductions we replaced
them with synthetic templates (\S\thinspace\ref{sec:iodine}). 

In addition to our program stars we obtained frequent observations of
two brighter stars (HD~209458 and HD~179949) that have known
low-amplitude velocity variations at the level of about 200 $\ms$ due
to orbiting substellar companions, and which we used as ``standards".
These stars were observed with the iodine cell.  Additionally we
observed a number of 10th-magnitude K stars repeatedly for a different
observing program, also with the iodine cell, which turned out to be
extremely useful as secondary standards for monitoring the stability
of the velocity zero point and verifying systematics for fainter stars.  
Identifiers and observational details
including exposure times for all objects discussed here are given in
Table~\ref{tab:stars}. All HIRES spectra were
bias-subtracted, flat-fielded, cleaned of cosmic rays, and extracted
using the MAKEE reduction package written by Tim Barlow
\citeyearpar{Barlow:02::}.  Wavelength solutions based on the Th-Ar
exposures were carried out with standard tasks in IRAF\footnote{IRAF
is distributed by the National Optical Astronomy Observatories, which
is operated by the Association of Universities for Research in
Astronomy, Inc., under contract with the National Science
Foundation.}. 

Radial velocities for OGLE-TR-3, OGLE-TR-33, and OGLE-TR-56 relied on
the Th-Ar calibration exposures for the wavelength reference. These
are referred to as `Th-Ar velocities'. For the other two targets we
used the iodine as the reference (`iodine velocities').  Th-Ar
velocities were derived by cross-correlation against synthetic
templates described below, using the IRAF task RVSAO
\citep{Kurtz:98::}. The final velocities are the weighted average of
all echelle orders. Formal errors were derived from the scatter of the
velocities determined from the different orders. These are typically
under $\sim$100 $\ms$, and do not include systematic components
discussed later, which we also estimate to be no larger than about 100
$\ms$. The velocities in the frame of the solar system barycenter and
their internal errors are listed in Table~\ref{tab:thar_rvs}. 

For OGLE-TR-10 and OGLE-TR-58 we used iodine reduction software
\citep[see, e.g.,][]{Korzennik:00::} with which we have typically
achieved precisions of $\sim$10 $\ms$ or better for a single
measurement on bright stars. For our much fainter OGLE targets the
limit to the final precision in the iodine velocities is signal to
noise, not systematics.  These iodine velocities are reported in
Table~\ref{tab:iodine_rvs}, and include barycentric corrections.  For
these targets and also for the standard stars the comparison between
the Th-Ar velocities and the iodine velocities allows us to estimate
the accuracy of the Th-Ar technique, as applied to the OGLE candidates
that were observed without iodine. 
	
\subsubsection{Stellar parameters and synthetic spectra}
\label{sec:templates}

The key physical properties of each of our program stars were
estimated from our high-resolution Keck spectra, by comparison with
calculated spectra.  These were computed from model atmospheres for
solar composition based on the ATLAS~9 and ATLAS~12 programs by
\cite{Kurucz:95::}, re-written in Fortran-90 (J.\ Lester, priv.\
comm.) and incorporating new routines for improved treatment of
contributions from various broadening mechanisms, as well as updated
and expanded opacities and line lists.  This code has been tested
extensively and performs very well for solar-type stars (F-K-type)
such as our targets.  All spectra for each OGLE target (both the
template spectra with no iodine, and the iodine exposures) were
shifted to a common reference frame and co-added. A comparison between
the observed co-added spectra and the synthetic spectra computed for a
range of stellar parameters was made in spectral regions unaffected by
the $I_2$ lines, and including a large number of metal absorption
lines of different ionization and excitation states as well as the
core and wings of the $\lambda$4861 H$\beta$ line. Effective
temperatures ($T_{\rm eff}$) were determined to an estimated accuracy
of about 100~K, surface gravities ($\log g$) to roughly $\pm$ 1~dex,
metallicity [Fe/H] to $\pm$ 0.3~dex, and projected rotational
velocities ($v \sin i$) to 2--3 $\kms$.  Synthetic templates with
these parameters (and degraded to the resolution of our Keck
observations) were used for all cross-correlations in order to obtain
radial velocities.  In Figure~\ref{fig:spectra} we show a section of
the co-added spectra for each of our stars in the H$\beta$ region, as
well as the best-fit synthetic spectrum and the values we derive for
the stellar parameters. In addition to comparing and fitting metal
lines of different ionization states and excitation potentials (mainly
for $T_{\rm eff}$ and [Fe/H]), we carefully fit the core vs.\ wings of
the Balmer H lines. Our approach is similar to that described by
\cite{FV:03::} regarding line broadening. 

The metal abundance for all our targets is consistent with the solar
value.  Other derived properties such as masses and radii for our
stars, used later in the analysis, were estimated with a stellar
evolution code described in detail elsewhere \citep{Cody:02::,S:03::}.
Since the distance to our targets is not known accurately, we used
$\log g$ as a proxy for luminosity. Evolutionary tracks are nearly
vertical in the $T_{\rm eff}$ vs.\ $\log g$ plane for our relatively
low-mass stars, so the fairly large uncertainties in the surface
gravity do not affect the inferred mass significantly for our
purposes. 

\section{Random and systematic errors in the Th-Ar radial
velocities: velocity standards and HIRES stability}
\label{sec:stability}

One of the key requirements for the measurement of highly precise
radial velocities is a reliable reference frame that enables one to
map small pixel shifts in the stellar lines into wavelength shifts.
To achieve $\ms$ precision with HIRES spectra, the shifts that need to
be measured are of the order of a few thousandths of a pixel. In the
classical approach that uses a comparison lamp, the accurate
measurement of velocities requires that the calibration spectrum
present a large number of suitable lines with which to compute a
wavelength solution, but most importantly, that there be no systematic
shifts between the stellar exposure and the lamp exposure(s). The
latter is the most serious limitation for HIRES, as we show in this
section, since the comparison spectra can only be taken before or
after the stellar spectrum. If shifts do occur, they can in principle
still be corrected for as long as they can be modeled accurately
(e.g., shifts that are linear with time).  The power of the iodine
technique comes from the fact that a very dense forest of $I_2$ lines
is imprinted \emph{simultaneously} with the stellar spectrum, thus
allowing shifts (and higher-order distortions) of the wavelength scale
to be measured accurately.  However, the use of the iodine cell
decreases the signal from the stellar spectra by a large factor (see
\S\thinspace\ref{sec:iodine}), which is a serious drawback for faint
objects.  Thus for most of the stars in our sample we were forced to
rely on the Th-Ar exposures for the wavelength reference.

To minimize shifts such as those mentioned above, each exposure of a
program star was bracketed by Th-Ar exposures taken immediately before
and after the stellar exposure.  In addition, velocity standards were
observed frequently to allow us to monitor residual instrumental
shifts.  Because these standard stars are much brighter than our
targets, they were observed with the iodine cell to take full
advantage of that technique. With our setup, approximately one third
of the echelle orders have $I_2$ lines and two thirds do not, so that
the orders unaffected by iodine can be used in the classical way for
the standard stars to obtain radial velocities using the Th-Ar lamp as
the reference. 

The standard stars were used for two main purposes: (1) To assess the
magnitude of the shifts that can occur between the Th-Ar spectra and
the stellar spectra on timescales similar to the exposures for the
OGLE targets (typically $\sim$30 minutes); and (2) to estimate the
intrinsic accuracy of the wavelength solutions based on the lamp
exposures and its impact on the velocities. 

An indication of instrumental shifts that can occur on timescales of
hours can be obtained by comparing the pixel locations of the emission
lines in the Th-Ar spectra taken during the night. For this, we
cross-correlated each lamp spectrum in pixel space against one of them
adopted as the reference, using the IRAF task FXCOR, and computed the
weighted average over all orders. These shifts are shown with filled
symbols in Figure~\ref{fig:thar1}, as a function of time.  Typical
errors for an individual shift are 0.005--0.009 pixel.  Additional
information on shifts throughout the night can be obtained from the
iodine spectrum imposed on the bright standard stars and on the K
stars. The high density of $I_2$ lines allows these shifts to be
determined very accurately relative to one of such spectra adopted as
the reference.  Aside from an arbitrary offset with respect to the
shifts measured from the Th-Ar exposures that is easily removed, the
iodine shifts are seen to follow a similar trend as the Th-Ar shifts
within each night (Figure~\ref{fig:thar1}, open symbols). It is clear
from the figure that there are significant shifts with time within any
given night and often with each new telescope pointing, although the
magnitude of these shifts is not entirely predictable. We point out
also that the shifts from night to night are not always meaningful, as
the position of the detector was occasionally adjusted manually (only
at the beginning of the night) to keep the Th-Ar lines within
$\sim$0.5 pixel of a nominal location on the CCD.  Shifts are seen to
be generally less than 0.5 pixel during a night, but even this
corresponds to a very significant shift in velocity space (1 $\kms$),
making it imperative to obtain Th-Ar exposure before as well as after
each stellar exposure. 

A key assumption in the Th-Ar analysis is that the shift between the
two comparison exposures is linear with time, so that interpolation to
the mid-time of the stellar exposure is essentially free from
systematics.  Demonstrating that this is actually the case by taking
long sequences of exposures under actual observing conditions would be
excessively time-consuming. However, given that these shifts seem
unavoidable, we can at least place useful limits on their residual
systematic effect by considering the largest shift that occurred
during our run.  In Figure~\ref{fig:thar2} we show the sequence of
Th-Ar exposures on each night, enlarged from Figure~\ref{fig:thar1},
and we indicate with vertical dotted lines the times of our
interleaved exposures of the OGLE candidates.  Consecutive exposures
of the same candidate are usually intended for cosmic ray removal, or
correspond to iodine-template pairs. The largest shift between Th-Ar
exposures occured on the last of our nights (for OGLE-TR-58), and
corresponds to 0.4 pixel over a 2-hour interval. This represents a
velocity shift of about 200 $\ms$ over 30 minutes, which is the
typical exposure time for our targets.  Since the interpolated
wavelength solutions we adopt are for the nominal center of each
exposure, we may assume that systematic errors will typically be half
of the total shift, or $\sim$100 $\ms$ (0.05 pixel) in this case. We
adopt this as a conservative estimate of our systematics for this run,
although shifts for other stars in our sample are probably smaller. 

The intrinsic accuracy of the Th-Ar technique was explored by
observing two bright standard stars (HD~209458 and HD~179949) multiple
times during the run, often 2 or 3 times during the same night.  In
addition we observed more than a dozen 10th-magnitude K stars for a
separate project.  All of these objects were observed with the iodine
cell, and also once without the iodine cell to provide a template.
Initially we did not anticipate reducing them in the classical manner,
so no Th-Ar exposures were obtained either before or after these
stellar exposures.  Nevertheless, they provide useful checks as we
describe below. 

Six of the K stars showed no iodine velocity variations at the level
of 10 $\ms$, and are therefore suitable as standards for our purposes
(see Table~\ref{tab:stars}).  Th-Ar velocities for these stars were
obtained from the non-iodine orders, by cross-correlation against the
corresponding non-iodine exposures used as templates. The wavelength
reference adopted was that provided by the first Th-Ar exposure of the
night (same for all stars on each night). The Th-Ar velocities for
each of the K stars during a given night contain an unknown systematic
error that has at least two components: an unknown instrumental shift
occuring between the time of the stellar (iodine) exposure and the
Th-Ar exposure at the beginning of the night, and another similar
shift associated with the wavelength calibration of the observed
template (which was not necessarily obtained on the same night). The
latter component is of course irrelevant when comparing velocities of
the same star, since the template is always the same.  We may thus
represent the measured Th-Ar velocity as
 \begin{equation}
\label{std:1::}
V_i(t_j) = \hat{V}_i(t_j) + W_i(t_j), \quad i=1,...,6, \quad j=1,...,4
 \end{equation} 
 where $V_i(t_j)$ is the velocity affected by systematics and
$\hat{V}_i(t_j)$ is the true velocity of the $i$-th star taken on the
$j$-th night. If we now use one of the stars (e.g., star number 1) as
the reference, we have
 \begin{equation} 
\label{std:2::}
\hat{V}_i(t_j) - \hat{V}_1(t_j) = V_i(t_j) - V_1(t_j) - [W_i(t_j) -
W_1(t_j)], \quad i=2,...,6, \quad j=1,...,4.
 \end{equation}
 Because the spectra of these K stars also have iodine lines in the
orders redward of about 5000~\AA, the differential shift $W_{i1}(t_j)
\equiv W_i(t_j) - W_1(t_j)$ can be determined very precisely (to
$\sim$10 $\ms$) using the iodine reduction software, and the
systematic errors mentioned above can effectively be removed resulting
in the Th-Ar velocity difference $\hat{V}_i(t_j) - \hat{V}_1(t_j)$.
This allows us to evaluate the intrinsic precision of the Th-Ar
technique, essentially free from systematics due to instrumental
shifts.  For this test we adopted HIP~1334 as star number 1, and
subtracted its velocity on each night from those of the remaining five
K dwarfs. In Figure~\ref{fig:kstarsys} we display these differences
$\hat{V}_i(t_j) - \hat{V}_1(t_j)$ after removing the average for each
star (to account for the difference in absolute velocities between
star $i$ and star 1).  The scatter of these differences, 49 $\ms$, is
a measure of the uncertainties intrinsic to the wavelength
calibrations in the classical Th-Ar technique, and the corresponding
share for a single measurement is about 35 $\ms$ (49
$\ms$/$\sqrt{2}$). This estimate reflects uncertainties that have to
do with rebinning in the cross-correlation process, template mismatch,
and other numerical uncertainties, and represents the absolute limit
of this technique for the HIRES instrument in the absence of other
shifts between the Th-Ar exposures and the stellar exposure, discussed
above. 

As a further test of the intrinsic uncertainties of the Th-Ar
technique, as well as of our ability to measure small changes in
velocity over our 4-night run, we used our spectroscopic observations
of HD~209458 and HD~179949, which are known to have low-mass orbiting
companions. Because the spectroscopic orbits of these stars are known
\citep{Mazeh:00::,Tinney:01::}, they allow us to study any biases that
may affect the determination of the velocity amplitudes. In addition,
their planetary companions are in the mass range that we expect to be
able to detect around the OGLE candidates, making them excellent test
cases. As in the analysis with the K stars, we measured the radial
velocities of HD~209458 and HD~179949 from the non-iodine orders using
the first Th-Ar exposure of the night as the wavelength reference and
the exposure without iodine as the template, and then subtracted the
Th-Ar velocities of HIP~1334 and applied the corrections
$W_{i1}(t_j)$.  These differences are shown in
Figure~\ref{fig:standards1} (filled circles). The intrinsic precision
of a single Th-Ar measurement, as estimated from the scatter for
HD~209458 and HD~179949, is 35 $\ms$ and 29 $\ms$, respectively (from
49 $\ms$/$\sqrt{2}$ and 41 $\ms$/$\sqrt{2}$), which are similar to our
result for the K stars. It is worth noting that this similarity is an
indication that photon statistics is not the dominant factor in the
estimate, since there is roughly a factor of two difference in S/N
between these two bright stars and the K dwarfs: the latter are
$\sim$3 mag fainter, and have exposures that are 5 times longer. 

A similar exercise with the two brighter standards but subtracting one
from the other is shown in Figure~\ref{fig:standards2}, where the
differential orbit is computed from the orbital elements of the two
stars.  This is analogous to Figure~2 by \cite{Konacki:03::}, except
that the corrections $W_{i1}(t_j)$ (where star 1 is HD~179949) have
now been applied to the differential Th-Ar velocities. The residual
scatter from the computed velocity difference (solid curve) is 69
$\ms$ (corresponding to 49 $\ms =$ 69 $\ms$/$\sqrt{2}$ for a single
velocity).  For comparison, the scatter without the corrections is 97
$\ms$, giving an error of 69 $\ms$ for a single velocity. 

From the above discussion we conclude that the intrinsic errors of the
Th-Ar technique with HIRES (unaffected by systematics) are smaller
than about 50 $\ms$ per measurement. An additional component to the
error for the OGLE stars comes from the lower signal-to-noise ratio of
their spectra compared to those of bright stars.  We estimate this
contribution to be around 60 $\ms$ (see next Section). Finally, the
main contributor to the overall error comes from the systematic shifts
(and their non-linearity) that we cannot control, between the lamp
exposures and the stellar exposures, which appear to occur on
timescales similar to the duration of the target exposures.  We have
conservatively estimated this to be $\sim$100 $\ms$, although in most
cases it should be less (see above). 

Thus our experiments show that, with appropriate observing protocols,
radial-velocity follow-up observations of objects as faint as $V =
16$--17, such as those from the OGLE-III survey, can achieve the
required precision of $\sim$100 $\ms$ to detect close-in extrasolar
giant planets using the classical Th-Ar technique. 
	
\section{Radial velocities with the iodine cell}
\label{sec:iodine}

As seen in the previous section, the fundamental limitation of HIRES
for obtaining very accurate radial velocities with the classical Th-Ar
technique is the unavoidable drifts in the location of the spectrum on
the detector that occur throughout the night, on timescales of hours
or less.  In particular, non-linearity of these shifts between the
time of the preceding and following comparison lamp exposures is
essentially impossible to account for. However, other instrumental
setups that rely on comparison lamps can overcome this limitation,
such as fiber-fed spectrographs that are designed to record Th-Ar
spectra interleaved with the stellar spectra during the same exposure.
As an example, the CORALIE spectrograph on the 1.2-m Swiss telescope
at the European Southern Observatory can regularly achieve errors of
10--15 $\ms$ or better with this technique \citep[see,
e.g.,][]{Udry:00::}.  In the case of HIRES on the Keck~I telescope, a
popular alternative is the use of the iodine gas absorption cell, as
described earlier, which can track shifts very accurately.  This
method is capable of reaching very high precision (and accuracy) of a
few $\ms$ \citep[e.g.,][]{Butler:96::} by carefully modeling the
instrumental profile of the spectrograph (often referred to as the
point-spread function, or PSF) explicitly, but it requires very high
S/N ratios (typically 100 or more) in order to achieve this.
Unfortunately, the absorption from the iodine cell effectively removes
up to one half or even two thirds of the light from the star,
depending on the spectral type, and it is generally considered that
the approach becomes impractical for stars fainter than about $V =
13$.  It is particularly onerous not only because of the long
exposures needed for faint stars with the iodine cell, but also
because of the even longer exposures required for the templates, which
must ideally be of much better quality so that their noise
characteristics do not completely undermine the entire procedure. 

However, if main goal is to limit the systematic errors (instrumental
shifts) that are at the level of $\sim$100 $\ms$ or so for HIRES, the
iodine-cell technique can still be used to advantage on faint stars
such as our OGLE candidates with a few modifications, as follows:

\noindent (1) The most important factor that limits the velocity
errors at the level of 10 $\ms$ for bright stars is the variability of
the PSF with time and even with position on the CCD. Thus modeling the
spectrograph PSF very carefully is a central feature of all iodine
reduction packages.  Given that the orbital periods of our OGLE
candidates are typically short (a few days), the expected velocity
variations due to planets similar to Jupiter are usually several
hundred $\ms$, rather than tens of $\ms$, and therefore we do not
require such high precision (which would be difficult to obtain for
faint stars, in any case, because of S/N limitations). This allows us
to simplify the iodine reduction software considerably by not modeling
the PSF at all, and only adopting an estimate for it. 

\noindent (2) The use of observed templates for each star in the
standard iodine reductions is another important feature that allows
the method to reach very high precision in the velocities, because it
removes any spectral mismatch in modeling the iodine spectra.  Since
obtaining a template exposure for each OGLE target with sufficient S/N
would be very expensive in terms of telescope time, we can avoid this
altogether by using calculated spectra instead. 

Two of our planet transit candidates were observed at Keck with the
iodine cell, anticipating that the analysis with that technique might
yield useful results: OGLE-TR-10 and OGLE-TR-58. In order to evaluate
the impact of the simplifications in the iodine reduction described
above, we have again made use of our spectra of HD~209458 and
HD~179949.  Synthetic templates for these stars were computed as
described in \S\thinspace\ref{sec:templates}, with physical parameters
as reported in the literature.  The resulting velocities for the two
stars are shown in Figure~\ref{fig:standards3} (solid symbols) and are
compared with the known orbits \citep{Mazeh:00::,Tinney:01::}.  The
standard deviation of the residuals is 26 $\ms$ for HD~209458 and 22
$\ms$ for HD~179949. These values are similar to the internal (formal)
errors reported by our software, suggesting that the procedure is
effective in removing systematic errors at this level. The
contribution from photon noise is expected to be minimal ($<$ 5--10
$\ms$) given the large S/N ratios of the spectra for these stars in
the iodine orders, which are in the range 100--300 per pixel. Thus,
most of the error presumably comes from the simplifications introduced
in the analysis (no PSF modeling, and template mismatch).  As an
illustration, the velocities computed using the observed templates
instead of synthetic templates are represented with open circles in
Figure~\ref{fig:standards3}.  Their scatter is 13 $\ms$ for HD~209458
and 10 $\ms$ for HD~179949, showing the level at which template
mismatch makes a difference. 

When the same procedure (using synthetic templates) is applied to
OGLE-TR-10 and OGLE-TR-58, which have much weaker spectra (S/N in the
range 15--25 in the iodine orders), the internal errors are typically
50--60 $\ms$ (see Table~\ref{tab:iodine_rvs}). Based on the results
for the brighter stars HD~209458 and HD~179949, and under the
assumption that the combined effects of systematics and other
simplifications in the method are similar for the OGLE stars and the
brighter stars (i.e., at the level of 20--30 $\ms$), we conclude that
photon noise is the dominant factor in the iodine velocity errors for
the OGLE targets.  Thus the errors listed in
Table~\ref{tab:iodine_rvs} provide realistic estimates of the total
uncertainties. These results are very encouraging, and suggest the
method may be very useful in the future for the spectroscopic
follow-up of other faint transiting planet candidates, from the OGLE
survey and other similar transit searches. 

\section{Analysis and discussion}
\label{sec:analysis}

High-precision radial velocity measurements such as those reported
above are crucial for establishing whether a star with a transit-like
light curve is orbited by a planetary companion, but they are
generally not sufficient because of the variety of and frequency with
which false positives occur
\citep[e.g.,][]{Queloz:01::,Yee:02::,Latham:03::}.  The lack of
detection of any significant velocity change at the level of 100 $\ms$
does not necessarily prove there is a substellar companion in orbit,
since the star may actually have a constant velocity, but be blended
with a much fainter eclipsing binary along the same line of sight that
is invisible in the spectrum and is the object responsible for the
shallow dips in brightness seen photometrically. The opposite case,
the detection of a small velocity variation, still does not constitute
absolute proof of a planetary companion and must be investigated
carefully to rule out other possible reasons for the apparent change.
For example, if the primary of a contaminating background eclipsing
binary is bright enough, its spectral lines would move back and forth
with the photometric period and could produce slight asymmetries in
the spectral lines of the main star that vary with phase. These, in
turn, might lead to completely spurious small-amplitude velocity
variations.  Therefore, in addition to measuring the Doppler shifts
for each of our OGLE targets, we have examined the profiles of the
spectral lines looking for strong asymmetries or asymmetries that vary
in phase with the photometric period. 

The cross-correlation analysis of the Keck spectra produces a
correlation function for each echelle order, which we used to derive
the Th-Ar velocities. The sum of these correlation functions is
representative of the shape of average line profile of the star.
Asymmetries for each spectrum were measured by computing the line
bisectors \citep[e.g.,][]{Gray:92::}, and are shown in
Figure~\ref{fig:bisectors} for OGLE-TR-3, OGLE-TR-10, OGLE-TR-56, and
OGLE-TR-58. Line asymmetries are present in all of these stars, but
they are not significantly larger than in the Sun (in which the
velocity span is typically a few hundred $\ms$). Given that the stars
themselves are quite similar to the Sun in effective temperature, we
conclude that these asymmetries are not unusual. Furthermore, they do
not show any obvious changes with photometric phase, which suggests
the Doppler shifts are not significantly affected. 

It is well known that chromospheric activity can induce spurious
changes in the velocity of a star, as well as in the photometry, that
are unrelated to the presence of an orbiting companion.  We have
examined the profiles of the \ion{Ca}{2}~H and K lines for each of our
targets for any signs of emission that might indicate high levels of
activity. Although the strength of the spectra at these wavelengths is
typically low, we see no obvious indications of emission in any of the
OGLE stars. 

Information from the light curves is also very helpful in ruling out
false positives. For example, the shape and duration of the transits
(ingress-egress, and flat portion) provides some constraint on whether
the companion can be a planet \citep[see][]{Seager:03::}, although it
is highly dependent on the quality of the light curve. The overall
shape of the light curve outside of eclipse also provides important
clues (see below). We have applied the prescriptions by
\cite{Seager:03::} to our OGLE targets to establish whether the nature
of the primary stars as inferred directly from the transit light
curves is consistent with the properties we derive from our spectra
(mainly the effective temperature and surface gravity).  Although
formally most stars pass the test, the uncertainties are such that in
general the results are inconclusive. 

In the following we examine the spectroscopic and photometric evidence
available for each of our candidates.
	
\subsection{OGLE-TR-3: a probable blend} 
\label{sec:ogle-3}

Very recently a claim has appeared in the literature that OGLE-TR-3
harbors a possible new transiting planet \citep{Dreizler:03::}.  These
authors obtained 10 radial velocity measurements over a period of 26
days in 2002, coincidentally the same month as our own observations,
using the UVES instrument on the 8.2-m VLT with a typical precision of
$\sim$100 $\ms$ per measurement. They examined three alternative
scenarios not involving a planet that might explain the available
observations. Two of these scenarios invoke blends with an eclipsing
binary along the same line of sight with $P = 1.19$~d \citep[the
photometric period determined by][]{Udalski:02a::}, and another
proposes that the true photometric period is actually twice the value
reported by \cite{Udalski:02a::}, and involves also a blend with an
eclipsing binary composed of two equal-size M stars.
\cite{Dreizler:03::} concluded that none of these scenarios is as
satisfactory as the planet hypothesis. 

Unfortunately our own velocity measurements of OGLE-TR-3 with the Keck
telescope do not allow us to test this claim, since we were only able
to observe the star on two nights, and the precision of the velocities
on one of them is rather low (see Table~\ref{tab:thar_rvs}).  Within
the errors we detect no significant change.  Nevertheless, we can
place other useful limits by examining line asymmetries and also
looking for the presence of a second set of lines in our spectra,
which might typically be expected in the case of a blend. As indicated
earlier, the spectral line bisectors shown in
Figure~\ref{fig:bisectors}a do not indicate any significant
asymmetries (the span of the bisectors is less than $\sim$200 $\ms$),
and they show no correlation with the phase in the orbit (i.e., the
photometric phase).  Similar conclusions on the lack of asymmetries
were reached by \cite{Dreizler:03::} using their own data.  Inspection
of our spectra using the two-dimensional correlation technique TODCOR
\citep{Zucker:94::} to search for signs of a second set of lines
(presumably the main star of the blended eclipsing binary) did not
reveal anything significant at the level of 5\% of the light of the
primary within $\pm$200 $\kms$ of its mean velocity.  This represents
a significantly stronger constraint than the 30\% limit reported by
\cite{Dreizler:03::}. 

Although the previous paragraph appears to support the
\cite{Dreizler:03::} claim in finding no evidence for a blend, careful
examination of their methods and observations lead us to conclude that
some of the evidence they presented in favor of a planet is not quite
correct, and a more complete analysis of all the data argues
\emph{against} the reality of a planet.

For the analysis of their radial velocity measurements
\cite{Dreizler:03::} adopted the photometric ephemeris for transits
published by \cite{Udalski:02a::}, which is $T = 2,\!452,\!060.22765 +
1.18990 \cdot E$, where $E$ is the number of cycles from the reference
epoch of transit.  An updated ephemeris based on a total of 14
recorded transits is available from the OGLE website\footnote{\tt
http://bulge.princeton.edu/$\sim$ogle/ogle3/transits/tab.html\label{foot:ogleeph}},
 \begin{equation}
\label{eq:neweph}
T = 2,\!452,\!060.24529 + 1.18917 \cdot E~,
\end{equation}
and is significantly better as can be seen easily from the tightness
of the folded photometric data, although it does not significantly
affect their conclusions --- or ours below\footnote{The original
ephemeris is based on the photometry from the first of the OGLE
observing seasons, June-October 2001. The OGLE data from the following
year folded with the same period show an offset of about 0.2 in phase,
which disappears when using the new period.}.

\cite{Dreizler:03::} state that they solved for the spectroscopic
orbital elements by holding the period fixed, and adjusting the
velocity offset ($\gamma$ in our notation, see eq. \ref{vel:01::}),
the velocity semi-amplitude $K$, \emph{and also a phase shift},
$\Delta\phi$. Given that the precision of their velocity measurements
is of the same order as the amplitude they attempted to fit, and that
the number of measurements is not large, allowing for a phase shift
instead of relying on the much better determined photometric epoch of
transit is highly inadvisable.  The available radial velocities (a few
of which have errors as large as 0.6 $\kms$) provide essentially no
useful constraint on the phase of OGLE-TR-3.  In describing their
velocity fit (illustrated in their Figure~7), the authors make the
statement that ``As expected, the derived radial velocity is indeed
zero at mid eclipse". While this is true, the slope of their radial
velocity curve at phase 0.0 is \emph{positive} instead of
\emph{negative}, clearly showing that this velocity curve is in fact
inconsistent with the transits.  By adopting the following model
 \begin{equation}
\label {vel:01::}
V_{\star} = \gamma - K\sin(2\pi[\phi(t)+\Delta\phi]),
\end{equation} 
 where $\phi(t) \in [0,1]$ is the orbital phase determined from
photometry (through equation~\ref{eq:neweph}), we are able to
reproduce their results for the amplitude and velocity offset using
their measured Doppler shifts.  However, the phase shift we obtain is
$\Delta\phi = 0.55 \pm 0.05$ rather than the value of 0.02 listed in
their Table~\ref{tab:thar_rvs} from their $\chi^2$ fit, possibly due
to the use by \cite{Dreizler:03::} of a positive sign in front of $K$
(which is in fact inconsistent with the photometric ephemeris).
Nevertheless, a plot of the velocities and computed curve from our
solution looks essentially the same as the one in their Figure~7,
confirming that for such fit there is indeed a shift of about half a
cycle compared to the photometric ephemeris, and this explains the
reversal of the slope at the primary eclipse. 

We believe the only sensible approach in a case such as this is to
adopt the phasing from the photometry (the updated ephemeris in
eq.~\ref{eq:neweph}), and to solve only for $\gamma$ and $K$.  This
solution results in a value of $K$ that is statistically
insignificant: $K = 0.100 \pm 0.061$ $\kms$. A graphical
representation of the data and this fit is shown in
Figure~\ref{fig:ogle3_1}a. 

This alone does not necessarily prove that there is not a planet in
orbit around the central star. It only shows that the star has no
significant velocity variation within the errors.  However, strong
hints \citep[also admitted by][]{Dreizler:03::} of a secondary eclipse
at phase 0.5 in the light curve of OGLE-TR-3 suggest
otherwise\footnote{This feature was overlooked in the initial
assessment of the photometry that led to the placement of OGLE-TR-3 on
our Keck observing program.  In retrospect, the star should not have
been considered as a good candidate.}. This is seen in
Figure~\ref{fig:ogle3_1}b, which is based on the original OGLE
photometry along with the ephemeris in eq.~\ref{eq:neweph}. More
objective ways of assessing the reality of a secondary eclipse (or
more generally, of the curvature outside of eclipse due to ellipsoidal
variations ---the telltale sign of a stellar companion) support these
indications.  For example, \cite{Drake:03::} carried out analytical
fits of the form $\cos 2\phi$ (sinusoidal modulation at twice the
orbital frequency) and concluded that the amplitude of the
out-of-eclipse variation in OGLE-TR-3 is significant ($1.49 \pm
0.37$~mmag). More recently \cite{Sirko:03::} improved on the analysis
and accounted also for correlations in the OGLE photometry that they
claimed led Drake to underestimate the errors.  They too concluded
that the amplitude of the ellipsoidal variation is significant ($2.67
\pm 0.67$~mmag), and also found a significant heating (reflection)
effect for OGLE-TR-3 of $2.58 \pm 0.73$~mmag, another indication of
the stellar nature of the companion.  On this basis they ruled it out
as a good candidate for planetary transits. 

Given the limits described at the beginning of this Section on
possible blend scenarios, we carried out extensive simulations in an
attempt to find configurations that are consistent with all the
observational evidence available for OGLE-TR-3. The techniques are
introduced and described in detail by Torres et al.\ (2003, in
preparation), and consist essentially of fitting the observed light
curve with a full-fledged eclipsing binary light curve solution
program constraining the stars to have physical properties derived
from theoretical isochrones and stellar evolution models. The
isochrones are the same for the eclipsing binary and the third star if
they form a physical triple system, or otherwise they can be
different. The properties of OGLE-TR-3 itself (effective temperature,
rotational velocity, estimate of the surface gravity) are constrained
to agree with our values for these quantities reported earlier.
Testable predictions can be made regarding the relative brightness of
the stars, the velocity amplitudes $K_1$ and $K_2$ of the eclipsing
binary, and the rotational velocities $v \sin i$ of its components
(assuming rotation is synchronized with orbital motion, presumably
valid for short periods).  We confirmed the claim by
\cite{Dreizler:03::} that a blend with an eclipsing pair of equal M
dwarfs having a period that is twice the reported period results in
eclipses that are too narrow (see their Figure~11). We also explored
configurations where a grazing eclipsing binary forms a physical
triple system with the main star we see in OGLE-TR-3, but in this case
the relative luminosity of the primary of the eclipsing pair is
predicted to be 40--50\% of the light of the third star in the
optical, which we can clearly rule out from our spectra (see above).
However, a situation in which the grazing eclipsing binary is in the
background and has its eclipses diluted by OGLE-TR-3 can produce an
acceptable fit (reduced $\chi^2 = 1.03$) and gives a predicted
relative optical luminosity for the primary star in the binary of the
order of 4\% or less, which is below the detection threshold in our
spectra.  The difficulty in detecting the spectral signature of such a
background star is compounded by the expected $v \sin i$ of $\sim$75
$\kms$, which smears out the spectral lines considerably.  The
contribution of the foreground star in this simulation (OGLE-TR-3
itself) is 89\% of the total light in the $I$ band, comparable to
estimates illustrated by \cite{Dreizler:03::} in their Figure~12. A
sample fit to the light curve for this particular blend case is shown
in Figure~\ref{fig:ogle3_2}. 

The a priori likelihood of such a blend scenario, in which the
eclipsing binary is aligned with OGLE-TR-3 to within about 1\arcsec\
(the resolution limit of the photometry) may not be very high, but it
cannot be ruled out completely, particularly in a very crowded field
such as toward the Galactic center. Given the evidence presented
above (i.e., signs of a secondary eclipse, and consistency with a
blend model having a grazing eclipsing binary in the background) along
with the lack of any significant velocity variations, we must conclude
as stated by \cite{Konacki:03::} that based on all the data available
OGLE-TR-3 is most likely \emph{not} orbited by a transiting planetary
companion.

\subsection{OGLE-TR-10: planetary companion or blend}
\label{sec:ogle-10}

Of the three spectra we obtained for OGLE-TR-10 using the iodine cell,
one is too weak to be usable and the other two show no significant
velocity variation within the errors, over an interval of two days
(Table~\ref{tab:iodine_rvs}). The line asymmetries are not unusually
large and show no obvious trend with phase, as seen in
Figure~\ref{fig:bisectors}b. Careful examination of the spectra reveal
no sign of another star brighter than about 5\% of the light of the
primary.  Also, the analysis by \cite{Sirko:03::} indicates that the
out-of-eclipse variations in the light curve are insignificant. 

In principle all of this evidence allows for the possibility of a
substellar companion. In fact, the formal velocity difference between
the two available measurements is consistent with the expected trend
from the most recent ephemeris ($P = 3.10140$~days; see
footnote~\ref{foot:ogleeph}), which is based on 7 or possibly 8
transits (see Figure~\ref{fig:ogle10_1}). As an exercise, a
determination of the velocity amplitude through
 \begin{equation} K =
[V_{\star}(\phi_1) - V_{\star}(\phi_2)]/(\sin\phi_2 - \sin\phi_1)
\end{equation} 
 results in the value $K = 100 \pm 43$ $\ms$ (where the error was
computed by propagating from the measurement errors using the above
equation), and a corresponding companion mass of $0.7 \pm 0.3$~M$_{\rm
Jup}$, well within the planetary regime. An analysis of the OGLE-III
light curve leads to an estimate of the radius of the companion of
$\sim1.3$~R$_{\rm Jup}$. 

The above is of course insufficient to demonstrate that the companion
is a planet, and a blend configuration could still account for all the
observations. In Figure~\ref{fig:ogle10_2} we show the results of our
blend simulation for this system, illustrating the fit for one such
scenario in which the star OGLE-TR-10 dilutes the light of a
background eclipsing binary comprised of an M1 main sequence star in
orbit around a G0 star, with the photometric period determined by the
OGLE team. In this case the orbit of the binary is exactly edge-on,
and the brightness of the G0 star in the optical is only 3\% of that
of OGLE-TR-10, which is below the detection threshold in our spectra.
OGLE-TR-10 contributes 94\% of the total light in the $I$ band. A very
shallow secondary eclipse only 0.003~mag deep is predicted in this
scenario, but is undetectable in the photometry available. We conclude
that while a planetary companion remains possible, a blend scenario is
also consistent with all observational constraints and may be more
likely given the significant crowding in the field toward the Galactic
center. Further observations are needed to resolve the issue. 

\subsection{OGLE-TR-33: a blend}
\label{sec:ogle-33}

Our analysis of this star based on 4 Keck spectra shows very clearly
that it is an example of a blend with an eclipsing binary (in this
case, a triple system), with obvious asymmetries in the spectral lines
that correlate with phase in the 1.95-day orbit.  A full description
of the observations and our results for OGLE-TR-33 will be reported
elsewhere (Torres et al.\ 2003, in preparation). 

\subsection{OGLE-TR-56: a Jupiter-size companion}
\label{sec:ogle-56}

The results from our radial velocities for this candidate were
reported by \cite{Konacki:03::}, who showed that it is orbited by a
planetary companion with a mass of $0.9 \pm 0.3$~M$_{\rm Jup}$ and a
radius of $1.30 \pm 0.15$~R$_{\rm Jup}$. The velocity measurements are
listed in Table~\ref{tab:thar_rvs}. As indicated in that paper, the
spectral line profiles show very little asymmetry and no discernible
change with the phase in the orbit, supporting the reality of the
velocity changes. The line bisectors for the individual spectra are
shown in Figure~\ref{fig:bisectors}c. Other than some contamination
(at the level of 10\%) from scattered moonlight, which was removed
using TODCOR as described by \cite{Konacki:03::}, we find no signs of
another star in the spectra at the level of $\sim$3\% or more of the
light of the primary. 

The photometric period of the system, $P = 1.21190$~days, appears
stable over the one-year interval spanned by the first and last
recorded transits. The 12 individual transits are shown in
Figure~\ref{fig:ogle56}. Four of these transit events are complete,
covering both the ingress and egress.  The analysis of the
out-of-eclipse variations by \cite{Sirko:03::} does not indicate any
significant ellipsoidal or heating effects, which would suggest a
stellar companion. 

Simulations as in the case of OGLE-TR-3 to test blend scenarios were
carried out and reported by \cite{Konacki:03::}. However, it is
important to note that in OGLE-TR-3 there is no significant velocity
variation, which leaves open the possibility that contamination from a
fainter eclipsing binary may be the underlying cause of the periodic
drops in brightness. For OGLE-TR-56, on the other hand, the velocity
changes are significant and cannot be explained by variable
asymmetries in the line profiles. This makes the blend scenario very
unlikely, and inconsistent with the available observations. We are
conducting further spectroscopic and photometric observations of this
system. 

\subsection{OGLE-TR-58: planetary companion or blend}
\label{sec:ogle-58}

A complication during the observation of this object arose at the
telescope due to the presence of two relatively bright stars at
separations of 1\farcs1 and 1\farcs7 on either side the target, more
or less aligned with the spectrograph slit.  This led us to modify the
usual observing protocols regarding corrections for atmospheric
dispersion.  The image rotator routinely used with HIRES normally
follows the parallactic angle so that atmospheric dispersion is
oriented along the slit.  For OGLE-TR-58 we set it at a different
angle roughly 90\arcdeg\ away from nominal in order to avoid having
the light from these two extra sources go down the slit along with
that of the target. This introduces the risk of spurious velocity
shifts whose magnitude will depend on the hour angle and elevation of
the star at the time of the observation.  Thus, the interpretation of
the two radial velocity measurements we obtained for this candidate
using the iodine cell (Table~\ref{tab:iodine_rvs}) is somewhat
problematic. The formal difference is not significant, although as in
the case of OGLE-TR-10 the trend with time is consistent with the
phasing from the OGLE-III photometry, for the reported period of $P =
4.34517$~days.  The velocity amplitude based on the two available
measurements is $K = 200 \pm 94$ $\ms$, and the estimated mass of the
companion is $1.6 \pm 0.8$~M$_{\rm Jup}$.  The light curve analysis
gives a tentative radius for the planet of $\sim1.6$~R$_{\rm Jup}$.
Spectral line asymmetries are not large (Figure~\ref{fig:bisectors}d). 

We note, however, that the photometry is also problematic for this
object, for several reasons: (1) The precision of the measurements is
somewhat worse than for other stars of similar brightness in the
OGLE-III survey, possibly due to the proximity of a bright star in the
field that is heavily saturated (Udalski 2002, priv.\ comm.); (2)
There appears to be a systematic offset in the light level outside of
eclipse of about 0.02~mag between the 2001 season and the following
season (see Figure~\ref{fig:ogle58_1}), possibly related to the bright
star mentioned above; (3) Only two clear dips in brightness have been
recorded for this star, during the first of the OGLE-III seasons, with
an interval of about 35 days. This renders the period rather
uncertain.  Although the reported period of 4.34~days fits best
(including what appears to be a third incomplete transit during the
2002 season), other periods cannot be entirely ruled out with the
existing data. Additional photometry
is needed to resolve these issues. 

Out-of-eclipse variations appear insignificant, according to the
analysis by \cite{Sirko:03::}, although this result could be affected
by the systematic offset noted above.  From the data available we
conclude that OGLE-TR-58 remains a good candidate for a transiting
planet, although a blend scenario is equally possible. 

\section{Final remarks}

Deep photometric searches for transiting extrasolar planets in the
field (OGLE-III, \citealt{Udalski:02a::}; EXPLORE,
\citealt{Mallen:03::}; and others) are beginning to uncover
significant numbers of candidates among faint stars ($V = 14$--20).
Similar searches in open clusters will soon follow suit (e.g., PISCES,
\citealt{Mochejska:02::}; STEPPS, \citealt{Gaudi:02::}). Our
spectroscopic follow-up of 5 of the OGLE-III candidates has presented
us with a sampling of the results that may be expected from other deep
surveys\footnote{Similar results have already been obtained recently
by \cite{Yee:02::}, for the EXPLORE project.}, and has highlighted the
challenges to be faced in confirming the planetary nature of any of
these objects. After considering all the spectroscopic and photometric
information together, two of our objects, OGLE-TR-3 and OGLE-TR-33,
show clear indications that the transit-like events are produced by
contamination from an eclipsing binary along the same line of sight.
On the other hand the case of OGLE-TR-56, reported earlier by
\cite{Konacki:03::}, is unlikely to be a blend and is the only
candidate with a measured radial velocity variation that is
significant. It harbors a Jupiter-size companion in a very tight orbit
with a period of only 1.2~days.  OGLE-TR-10 and OGLE-TR-58 remain good
candidates for having a planetary companion similar to Jupiter, but
the data available so far also allow for the possibility that they are
blends. 

Our high-resolution spectroscopic observations of a number of standard
stars have enabled us to characterize the stability of the HIRES
instrument on Keck~I quite accurately. We have shown that the main
contribution to the errors of velocities determined by the classical
Th-Ar technique appears to be systematics at the level of $\leq$100
$\ms$, and is due to shifts of the spectrum on the detector beyond the
observer's control. Numerical errors intrinsic to the technique are
smaller ($\leq$50 $\ms$). Even for fainter stars such as the OGLE
candidates, systematics dominate in most cases. This indicates that
with proper care it is possible to measure Doppler shifts of faint
stars to about 100 $\ms$. We view this as a very encouraging result
for future follow-up campaigns with this instrumentation. It is worth
pointing out that, while precisions of a few $\ms$ may never be
achievable for faint stars with current telescopes and instrumentation
(and thus small planets in large orbits are beyond reach), a
significant portion of parameter space remains in which variations of
a few hundred $\ms$ are still detectable in stars as faint as $V =
17$, namely, planets similar to Jupiter with relatively short orbital
periods of a few days. Precisions of $\sim$100 $\ms$ such as we have
demonstrated here are sufficient to detect such objects. 

We have also shown that even better precision (and accuracy) of 50--60
$\ms$ can be obtained for faint stars by using the iodine gas
absorption cell on HIRES, with two modifications to the standard
procedures that make the technique as efficient as the classical Th-Ar
method. The first is that detailed PSF modeling is not required since
the photon noise for faint stars is much larger than the level at
which that modeling makes any difference. The second is that the
modeling works sufficiently well with synthetic templates instead of
observed templates, thus obviating the need to spend precious
telescope time obtaining high S/N spectra of each candidate without
the iodine cell. 
	
\acknowledgments

We are grateful to A.\ Udalski and the OGLE team for their many
generous contributions to this project. We also thank K.\ Stanek for
his continuous encouragement, and D.\ Mink for help with software
issues. D.S. thanks S.\ Korzennik for his invaluable help in
understanding precision radial velocities. The data presented herein
were obtained at the W.\ M.\ Keck Observatory, which is operated as a
scientific partnership among the California Institute of Technology,
the University of California and the National Aeronautics and Space
Administration. The Observatory was made possible by the generous
financial support of the W.\ M.\ Keck Foundation. M.K.\ gratefully
acknowledges the support of NASA through the Michelson fellowship
program and partial support by the Polish Committee for Scientific
Research, Grant No.~2P03D~001~22., and G.T.\ acknowledges support from
NASA's Kepler mission. S.J.\ thanks the Miller Institute for Basic
Research in Science at UC Berkeley for support through a research
fellowship. This research has made use of the SIMBAD database,
operated at CDS, Strasbourg, France, and of NASA's Astrophysics Data
System Abstract Service.

\clearpage

%
%

\figcaption[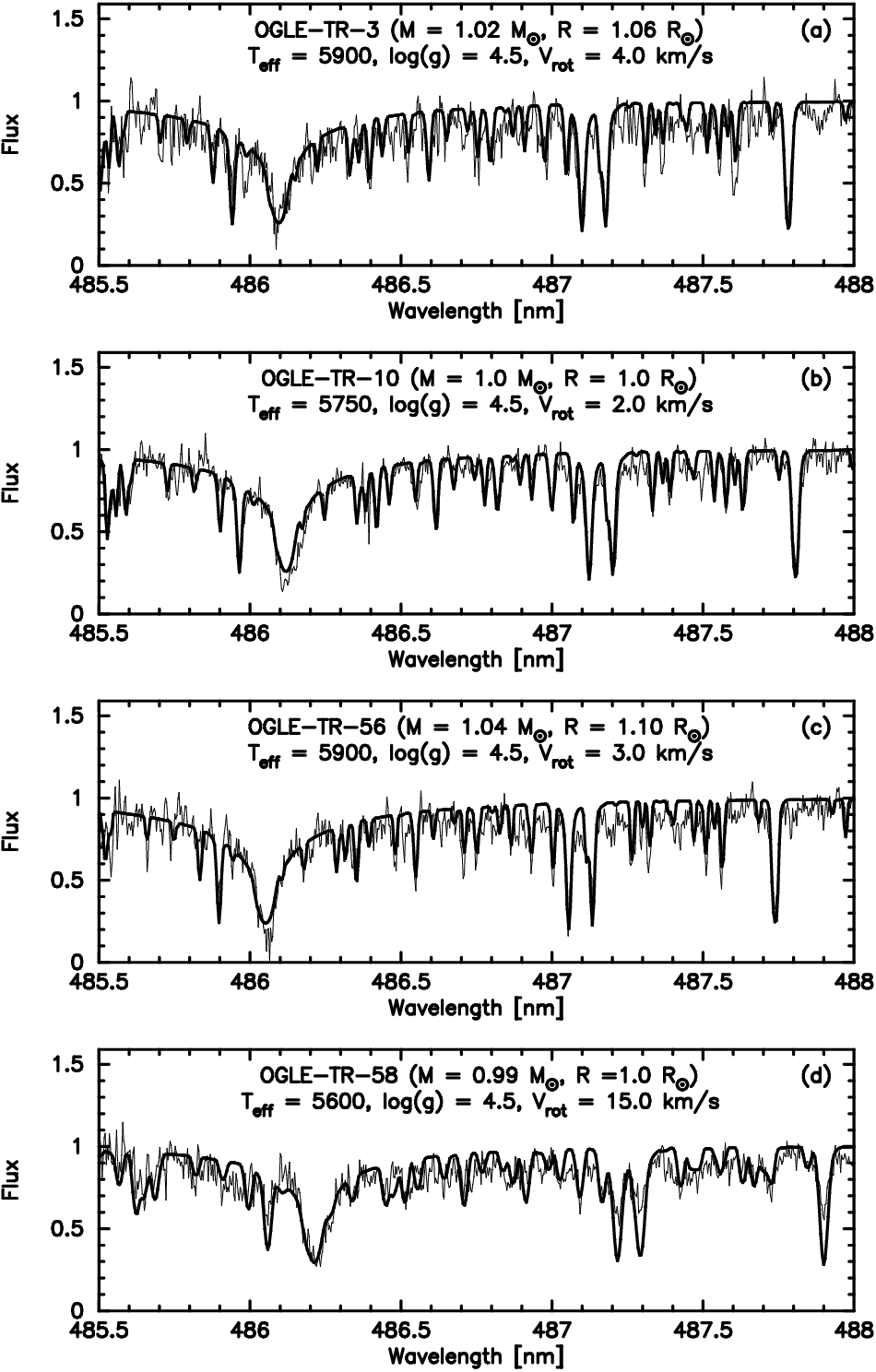]{Portion of the observed (co-added) spectra around the
H$\beta$ line for four of our OGLE candidates, with the best-fit
synthetic spectra superimposed (smooth solid line). The derived
stellar parameters are indicated.
\label{fig:spectra}}

\figcaption[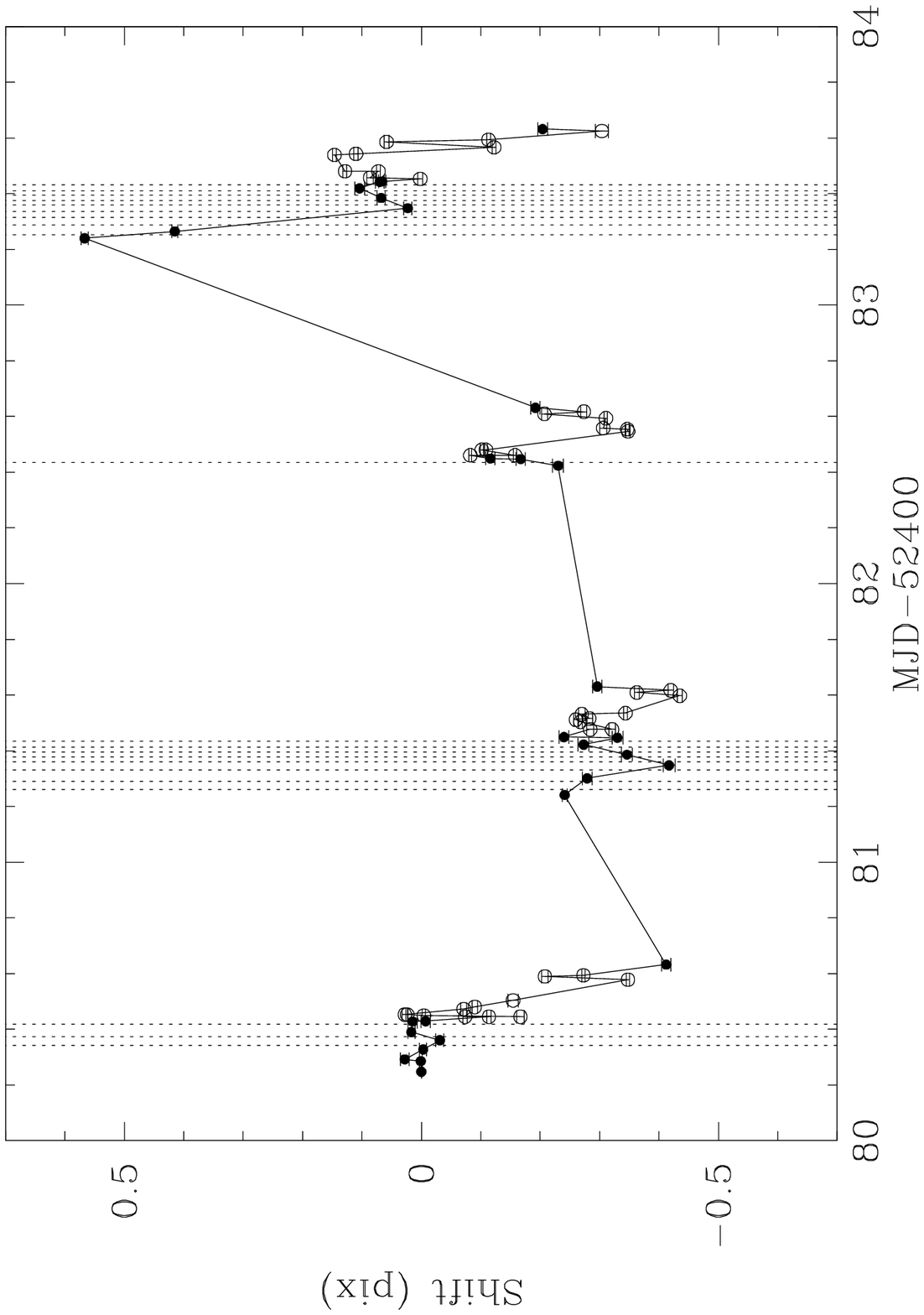]{Record of the shifts (in pixels) measured on the HIRES
detector as a function of time during our Keck run.  Filled circles
represent the shifts of each of the Th-Ar exposures compared to the
first such frame, and open symbols correspond to shifts measured on
stellar exposures of standard stars and K stars with the iodine cell,
also compared to the first such exposure (see text). An offset has
been applied to the iodine shifts to bring them into agreement with
the Th-Ar shifts.  Error bars for each shift are plotted, but are
smaller than the symbol sizes.  Vertical dotted lines show the times
of the exposures for the OGLE candidates.\label{fig:thar1}}

\figcaption[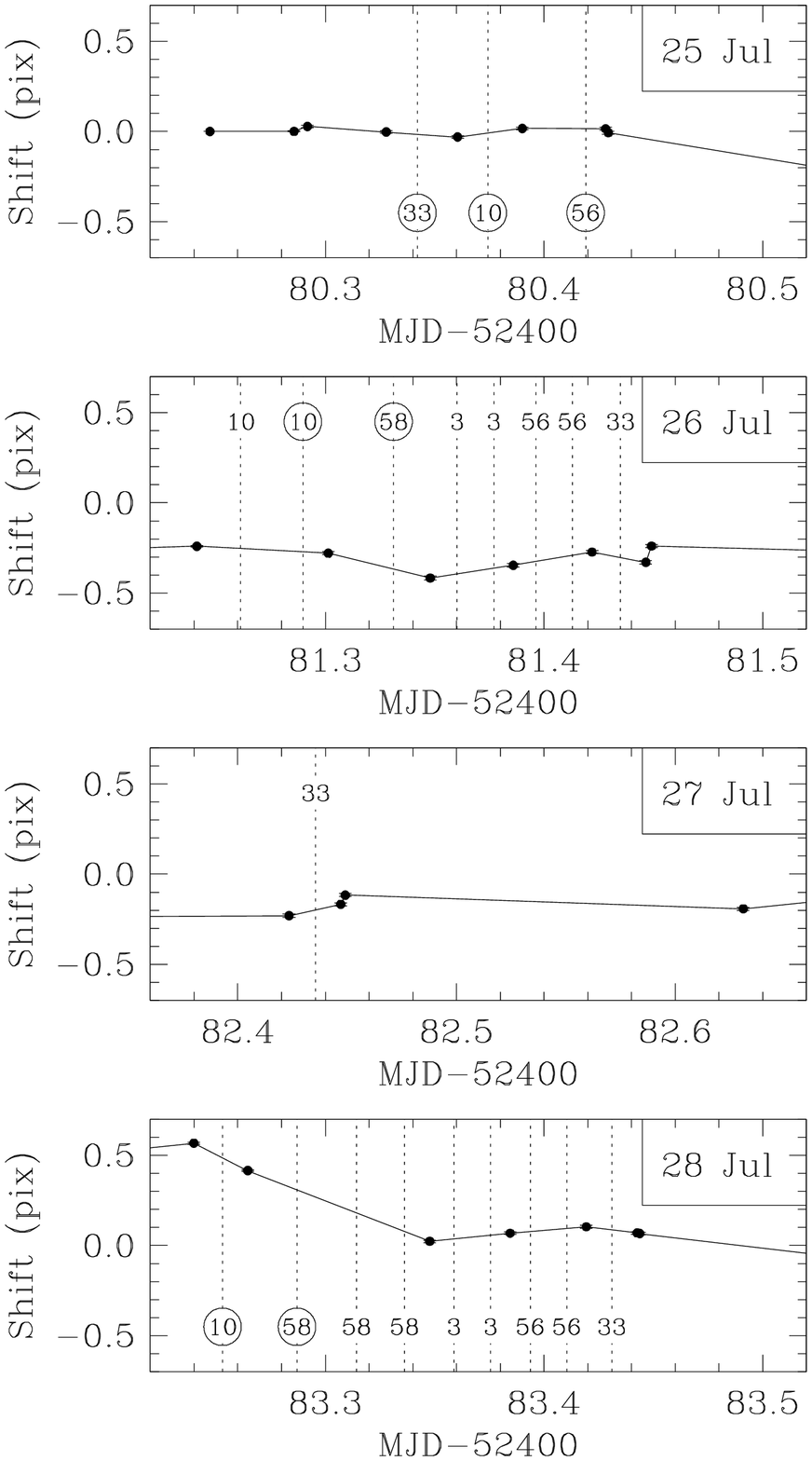]{Close-up of Figure~\ref{fig:thar1} showing the shifts as
measured from the Th-Ar exposures separately for each night. The times
of each OGLE observation are shown with vertical dotted lines, and
labeled with the number of the candidate (e.g., `33' for OGLE-TR-33).
Circled numbers indicate iodine exposures.\label{fig:thar2}}

\figcaption[f4.esp]{Differential radial velocities of the five `constant' K
dwarfs HIP~1078, HIP~2102, HIP~3232, HIP~117177, and HIP~117846
determined with the Th-Ar technique, and referred to the velocities of
HIP~1334. Systematic effects have been removed using the iodine (see
text), and the average for each star has been subtracted. The standard
deviation of these differences is 49 $\ms$.\label{fig:kstarsys}}

\figcaption[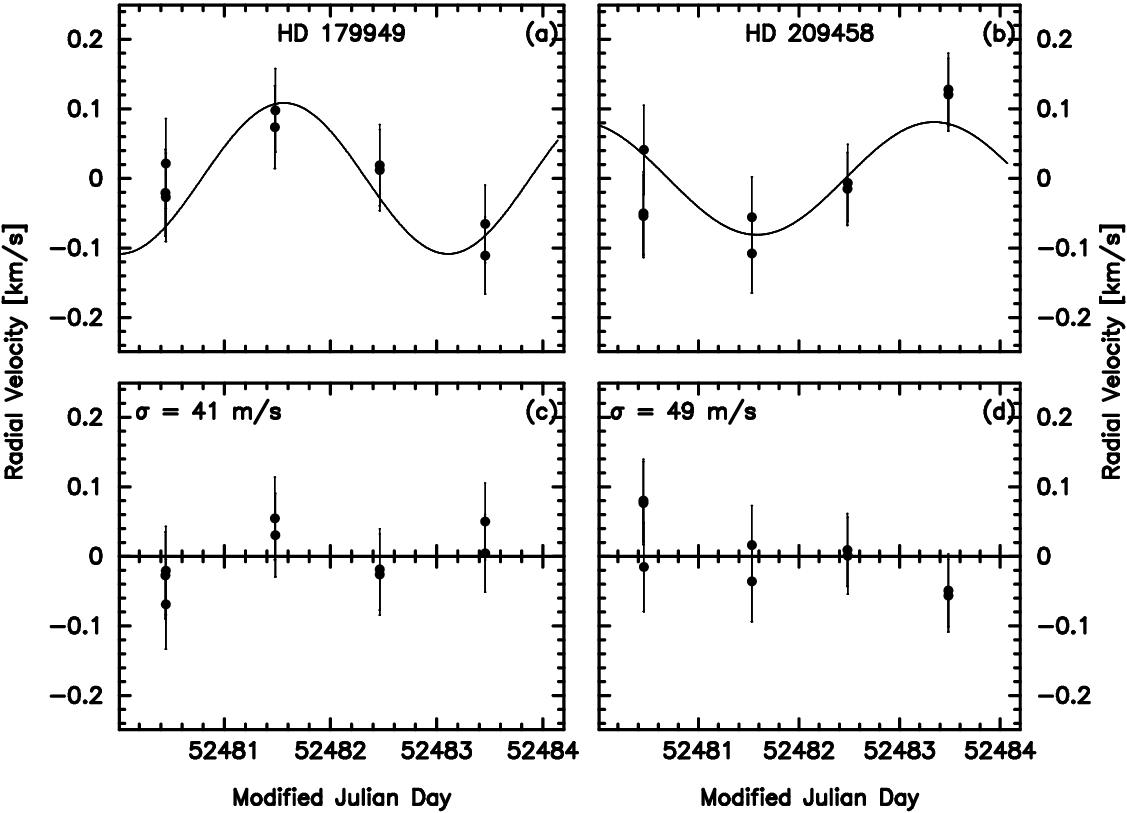]{Differential radial velocities of HD~179949 (a) and
HD~209458 (b) relative to HIP~1334, determined with the Th-Ar
technique and corrected for systematics (see text).
The solid lines
represent the orbits due to the known planetary companions of these
stars \citep{Mazeh:00::,Tinney:01::}. Residuals of the Th-Ar
velocities from these orbits give standard deviations of 41 $\ms$ for
HD~179949 (c) and 49 $\ms$ for HD~209458 (d).\label{fig:standards1}}

\figcaption[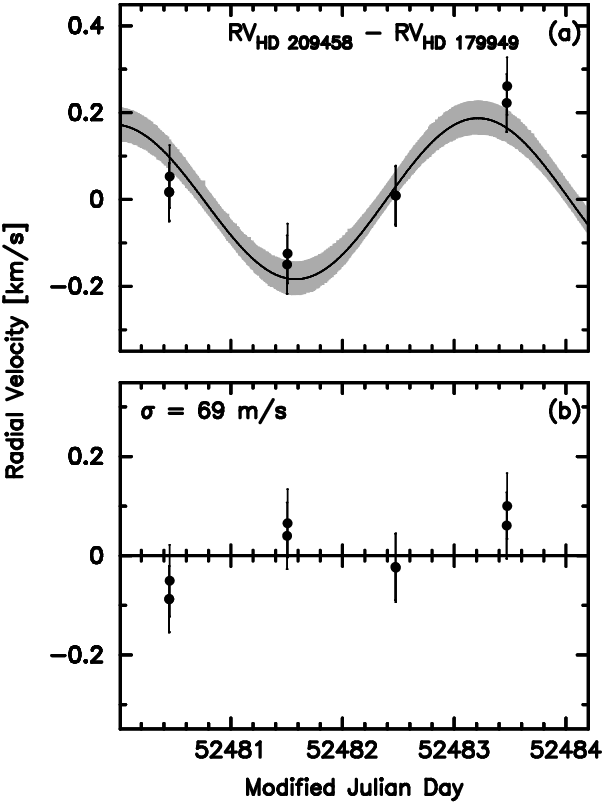]{Differential radial velocities (HD~209458 minus HD~179949)
from the Th-Ar technique, corrected for instrumental
shifts using the iodine orders (a). This is analogous to
Figure~\ref{fig:standards1}b, except for the use of HD~179949 as star
1 instead of HIP~1334 (see text). The differential orbit computed from the
elements published by \cite{Mazeh:00::} and \cite{Tinney:01::} are
indicated by the solid line. The shaded area represents the
uncertainty introduced by errors in the orbital elements. After
subtracting the orbital variations, the standard deviation of the
differential Th-Ar velocities is 69 $\ms$ (b).\label{fig:standards2}}

\figcaption[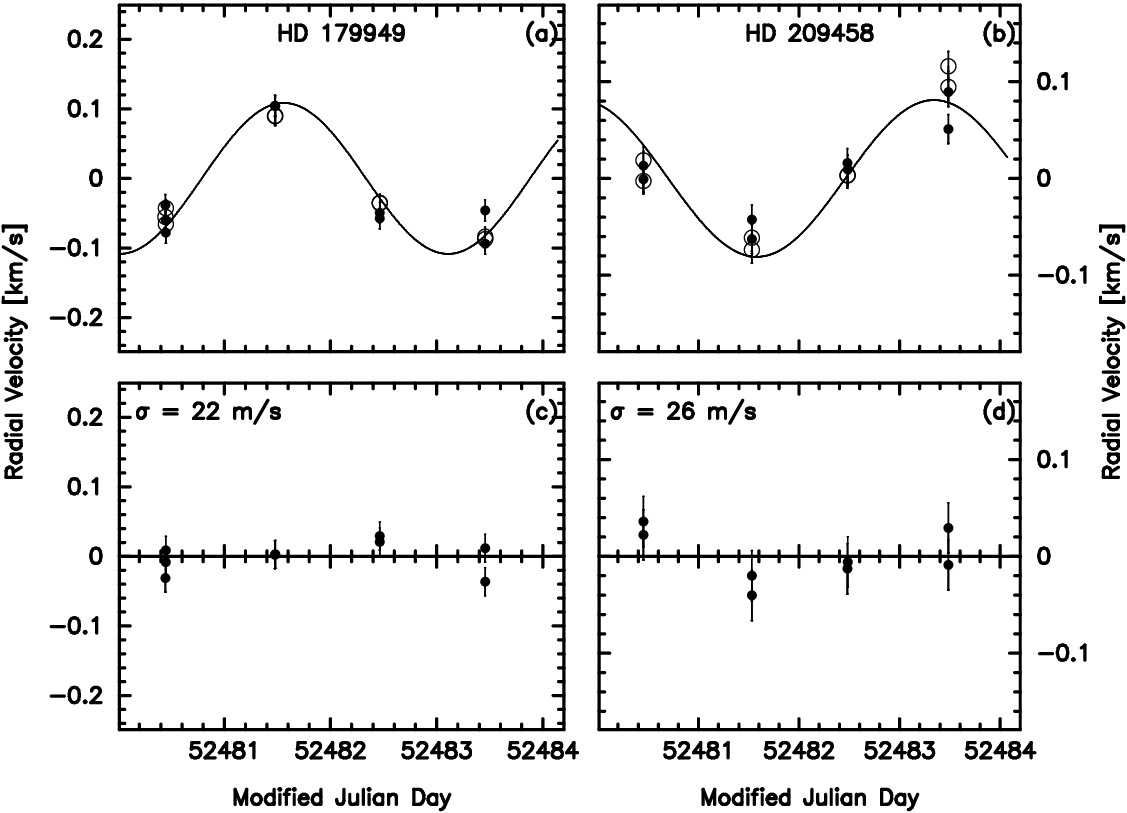]{The radial velocities of HD~179949 (a) and HD~209458 (b)
determined with the $I_2$ cell using synthetic templates.  The solid
line represents the calculated velocity differences
\citep{Mazeh:00::,Tinney:01::}. After subtracting the orbital
variations, the standard deviation of the $I_2$ velocities is 22 $\ms$
for HD~179949 (c) and 26 $\ms$ for HD~209458 (d). As a test, we
derived the iodine velocities also using observed templates (open
circles). The scatter in this case is smaller (see text). 
\label{fig:standards3}}

\figcaption[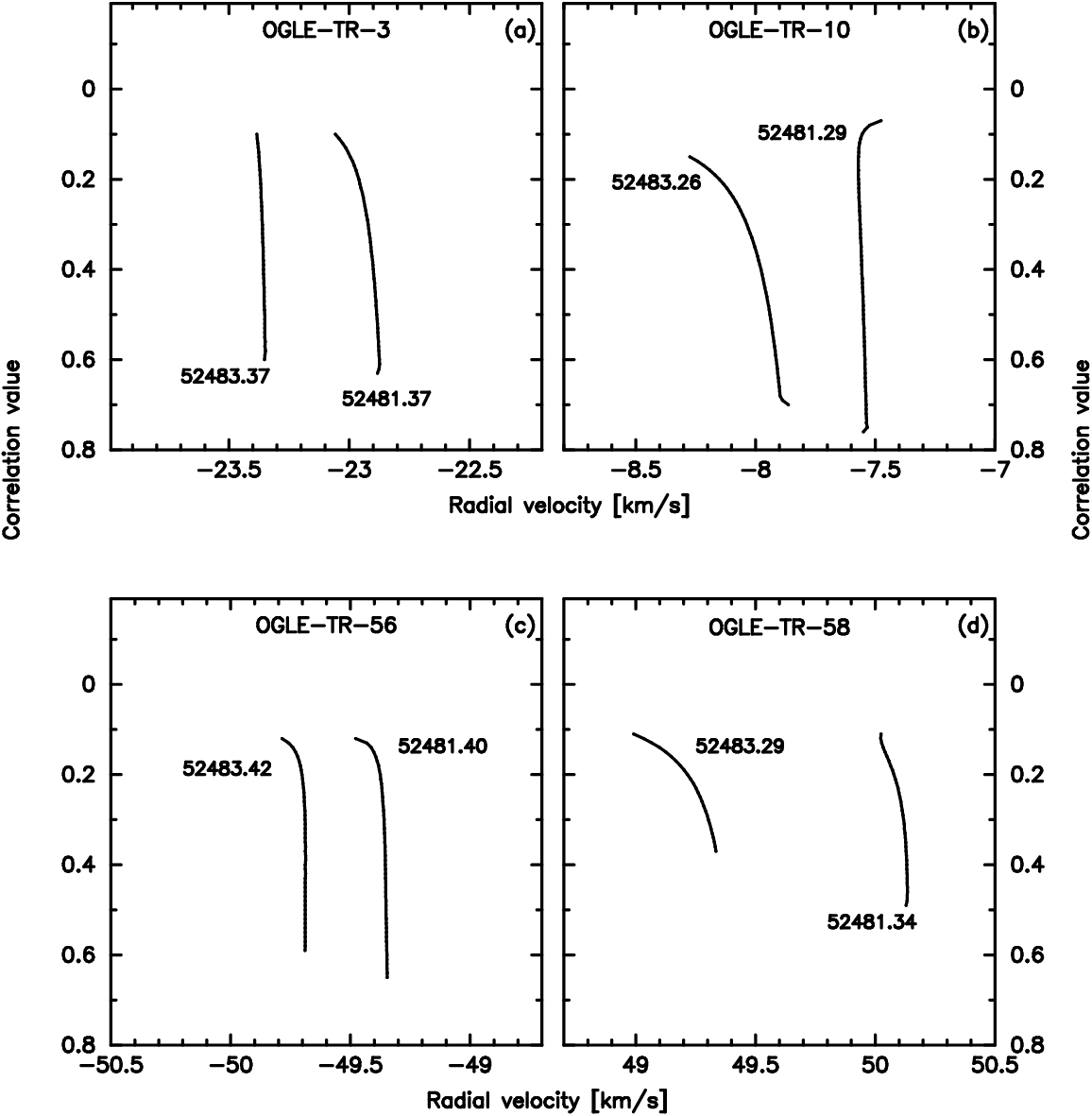]{Spectral line bisectors for OGLE-TR-3, OGLE-TR-10,
OGLE-TR-56 (excluding a weak exposure), and OGLE-TR-58, labeled with
the date (MJD) of each observation.  There are no asymmetries
significantly larger than in the Sun (see text), or any correlations
with photometric phase.  \label{fig:bisectors}}

\figcaption[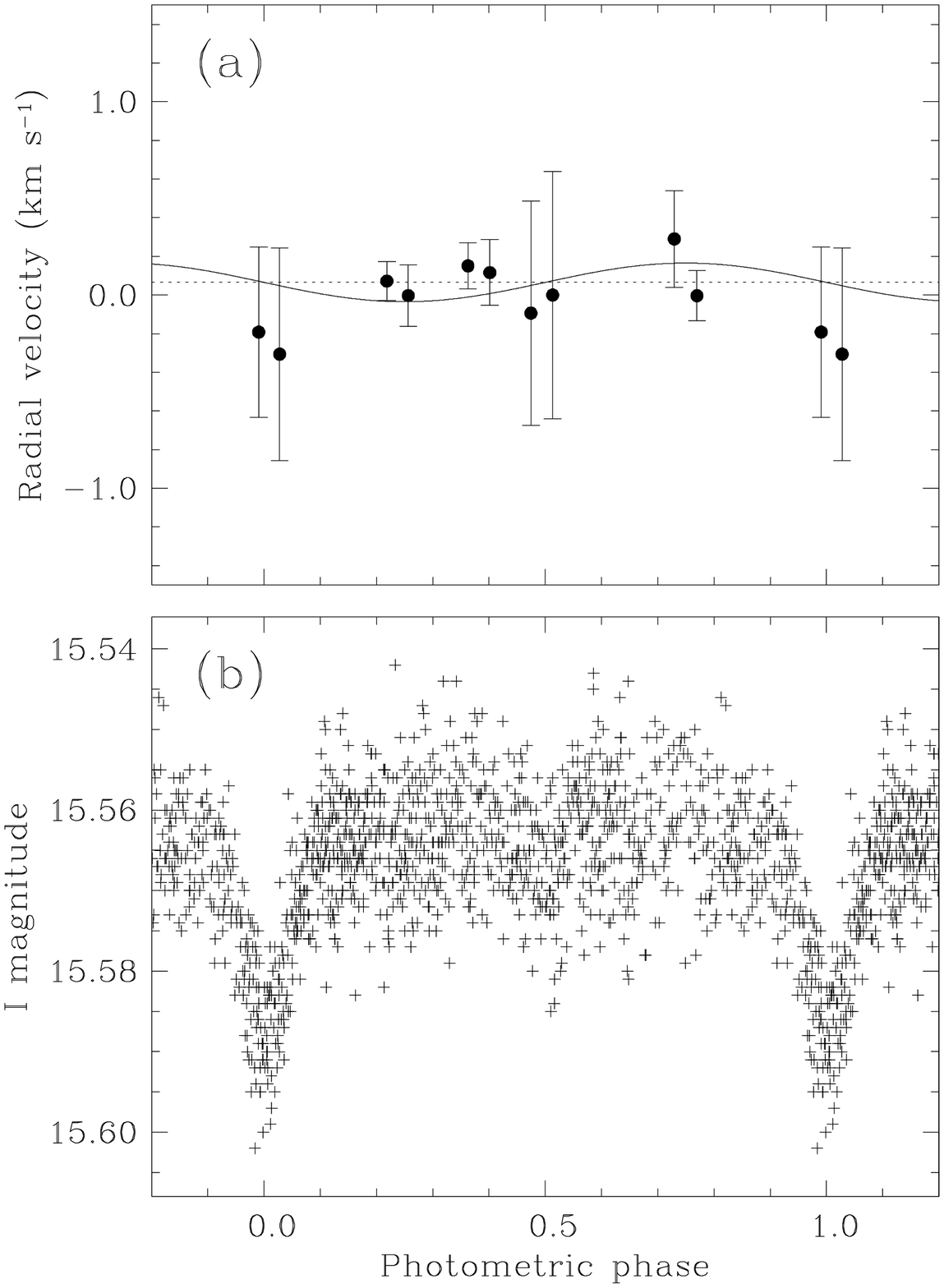]{(a) Radial velocity measurements of OGLE-TR-3 by
\cite{Dreizler:03::} along with our orbital fit of those data,
adopting the ephemeris in eq.~\ref{eq:neweph}. The velocity amplitude
is insignificant: $K = 0.100 \pm 0.061$ $\kms$; (b) OGLE photometry
phased with the same ephemeris. The hint of a secondary eclipse at
phase 0.5 and the out-of-eclipse variations are very suggestive (see
text).\label{fig:ogle3_1}}
 
\figcaption[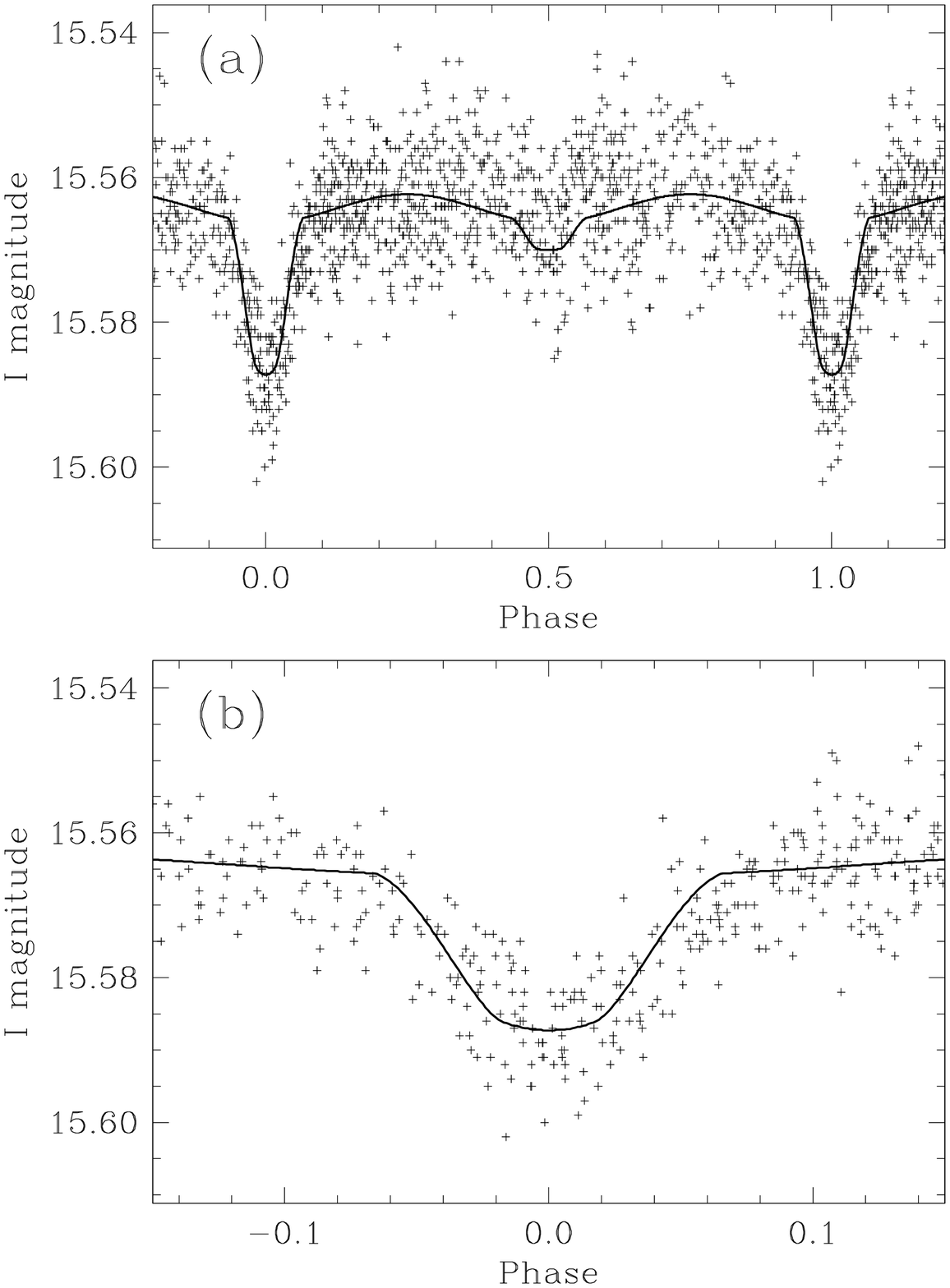]{Light curve fit resulting from a blend model with a grazing
eclipsing binary in the background of OGLE-TR-3. The physical
properties of the latter star (effective temperature, rotational
velocity, surface gravity) are constrained by the estimates from our
Keck spectra. The eclipsing binary in this model consists of a mid-K
star orbiting a mid-F star, with an inclination angle of 82\arcdeg.
(a) Full light curve; (b) Enlargement of primary
eclipse.\label{fig:ogle3_2}}

\figcaption[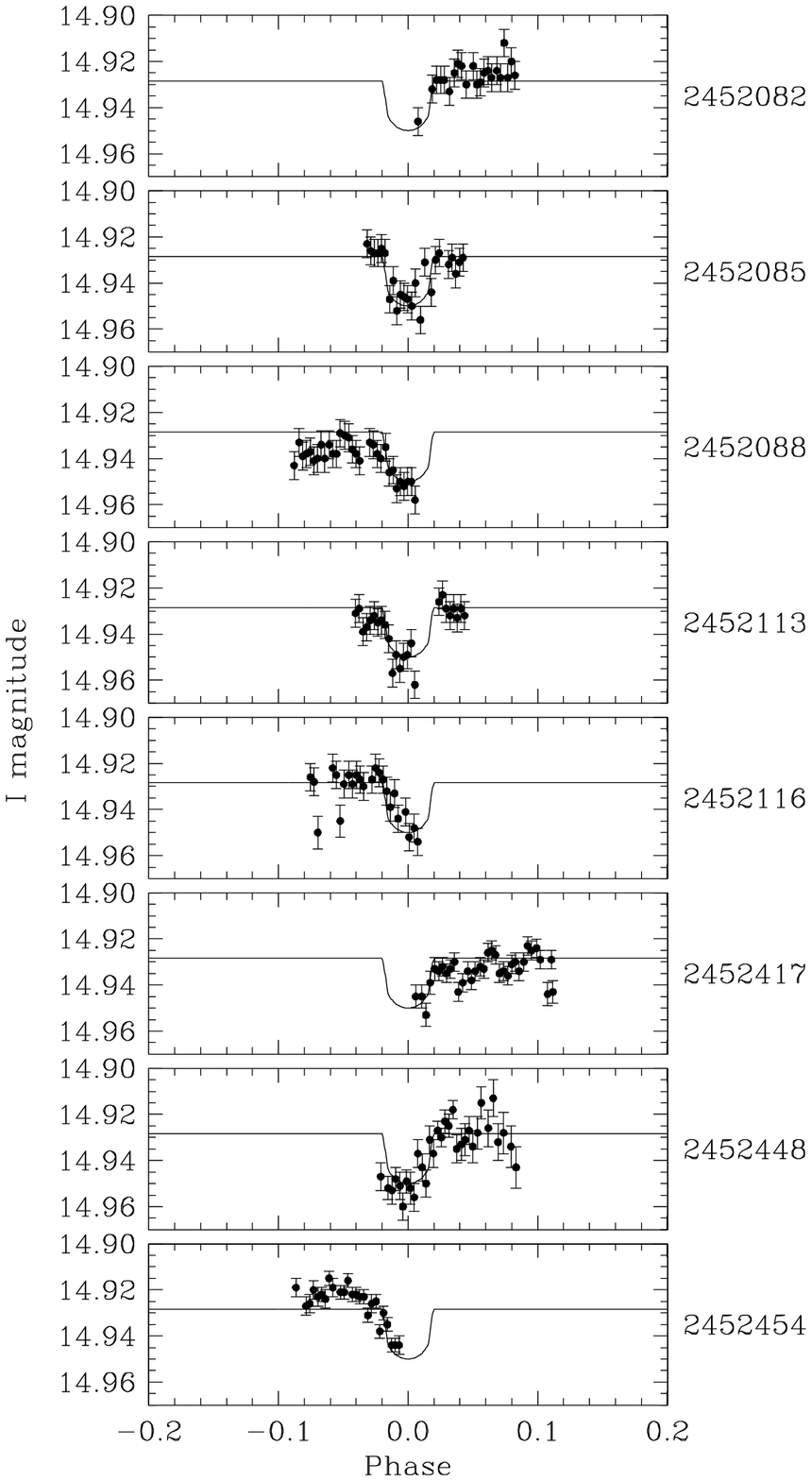]{Individual transits of OGLE-TR-10 recorded in the OGLE-III
photometry during the 2001 and 2002 observing seasons. Julian dates
for the corresponding night are indicate for each panel.  A tentative
fitted model is shown for reference.\label{fig:ogle10_1}}

\figcaption[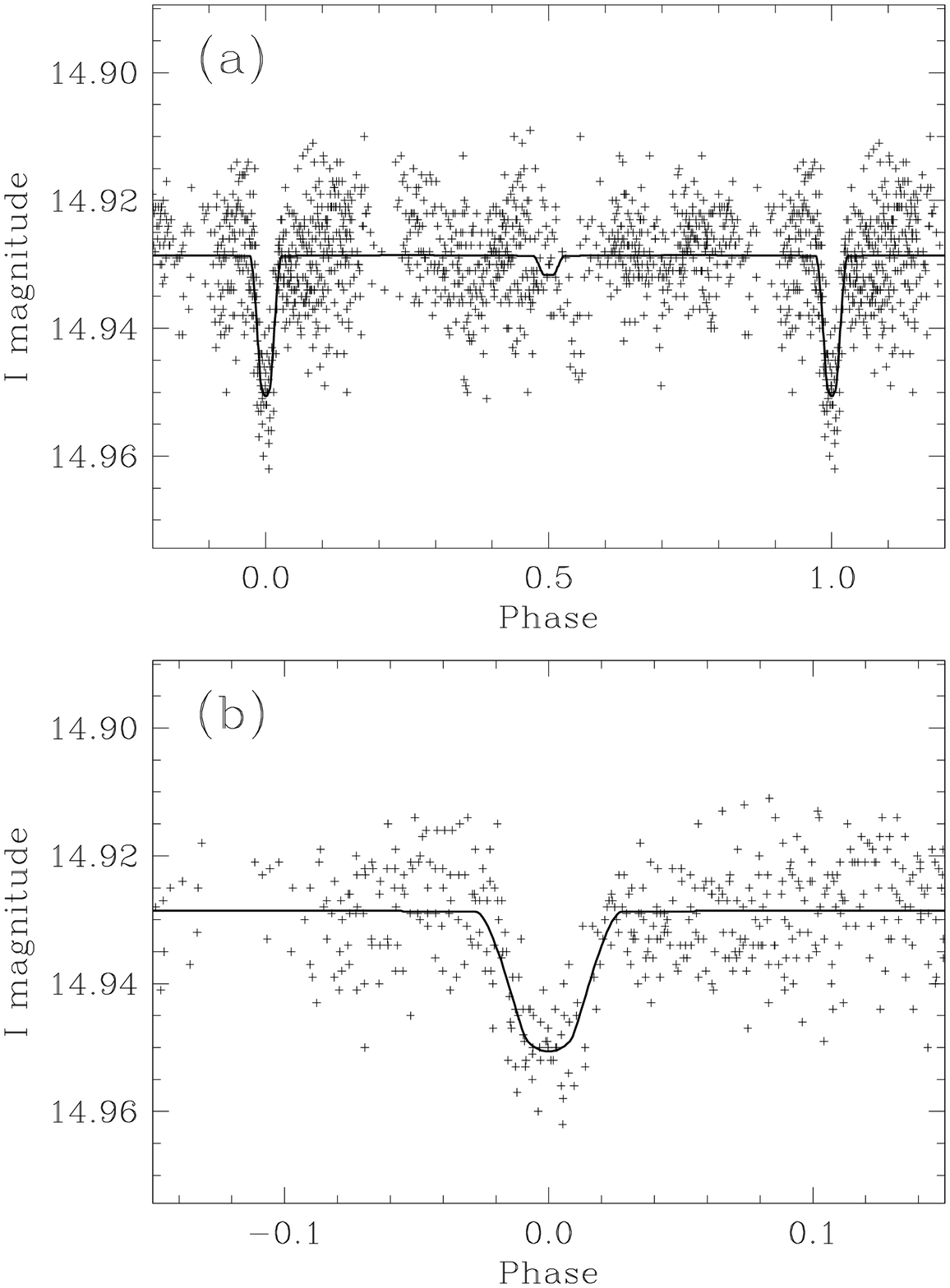]{Light curve fit resulting from a blend model with an edge-on
eclipsing binary in the background of OGLE-TR-10. The physical
properties of the latter star (effective temperature, rotational
velocity, surface gravity) are constrained by the estimates from our
Keck spectra. The eclipsing binary in this model consists of an
early-type M star orbiting a G0 star.  (a) Full light curve; (b)
Enlargement of primary eclipse.\label{fig:ogle10_2}}

\figcaption[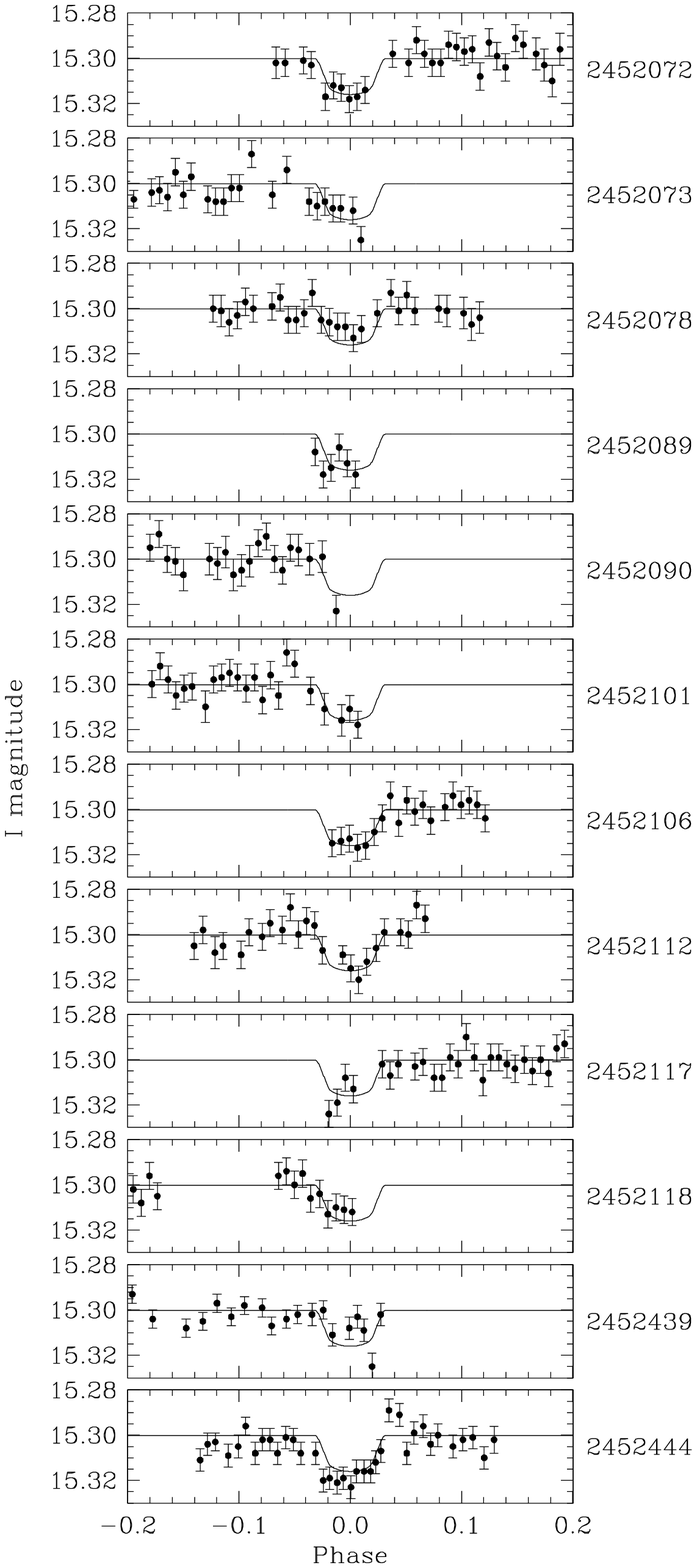]{Individual transits of OGLE-TR-56 recorded in the OGLE-III
photometry during the 2001 and 2002 observing seasons. Julian dates
for the corresponding night are indicate for each panel. The fitted
model from \cite{Konacki:03::} is shown for
reference.\label{fig:ogle56}}

\figcaption[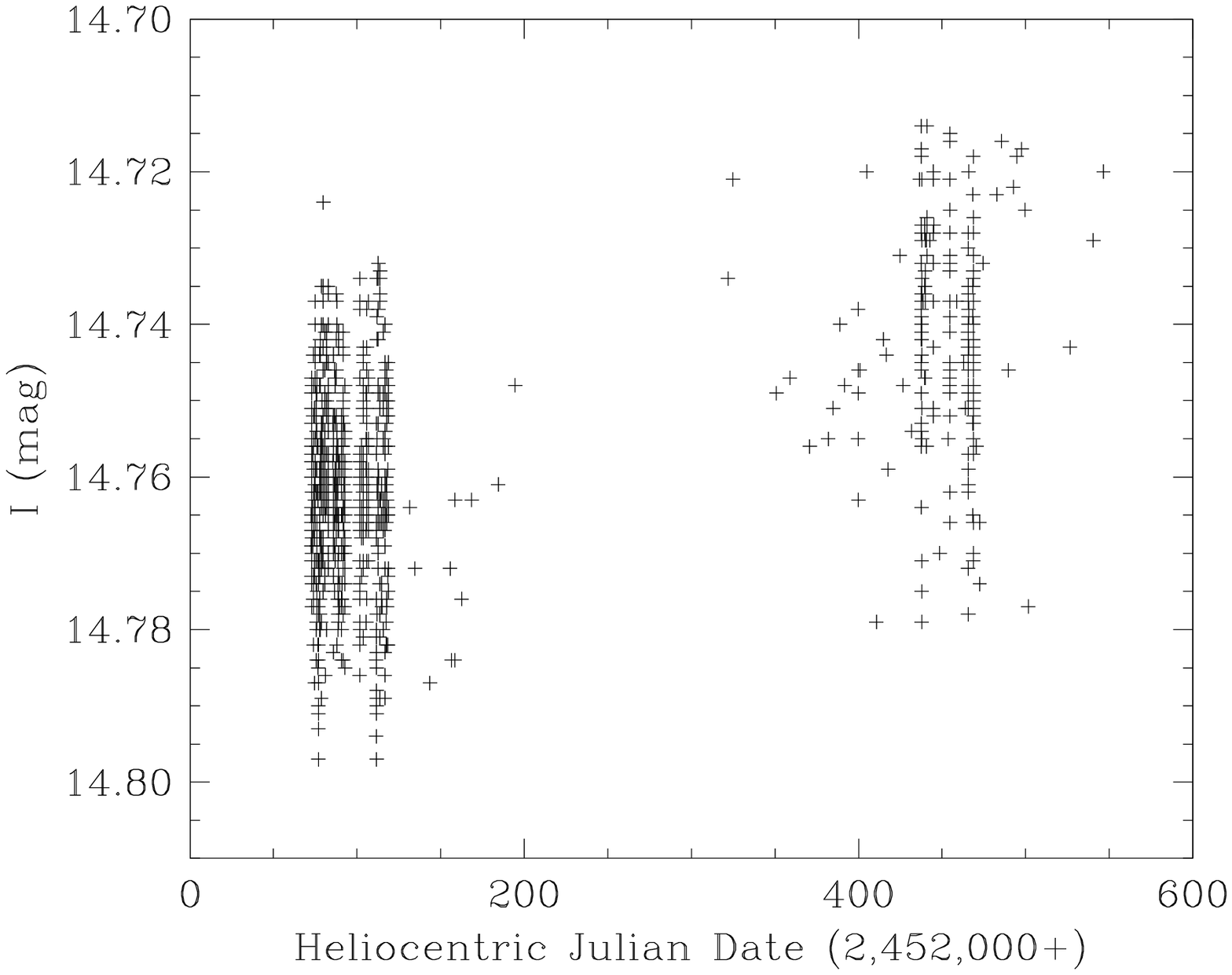]{OGLE-III photometry for the candidate OGLE-TR-58, as a
function of time. A systematic offset of about 0.02~mag between the
two observing seasons is apparent, and is possibly related to the
presence of a bright saturated star nearby.\label{fig:ogle58_1}}

\clearpage

%
%

\begin{figure}
\figurenum{1}
\hskip 0.9in \includegraphics[scale=0.75]{f1.eps}
 \caption{}
 \end{figure}

%
%

\begin{figure}
\figurenum{2}
\hskip -1.0in\includegraphics[angle=-90,scale=0.8]{f2.eps}
\vskip -0.5in
 \caption{}
 \end{figure}

%
%

\begin{figure}
\figurenum{3}
\epsscale{0.9}
\plotone{f3.eps}
\vskip 1.2in
 \caption{}
 \end{figure}

%
%

\begin{figure}
\figurenum{4}
\epsscale{0.85}
\plotone{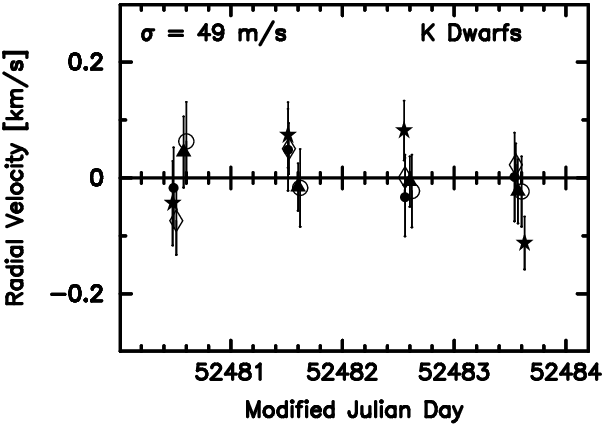}
 \caption{}
 \end{figure}

%
%

\begin{figure}
\figurenum{5}
\epsscale{0.85}
\plotone{f5.eps}
 \caption{}
 \end{figure}

%
%

\begin{figure}
\figurenum{6}
\epsscale{0.5}
\plotone{f6.eps}
 \caption{}
 \end{figure}

%
%

\begin{figure}
\figurenum{7}
\epsscale{0.8}
\plotone{f7.eps}
 \caption{}
 \end{figure}

%
%

\begin{figure}
\figurenum{8}
\epsscale{0.8}
\plotone{f8.eps}
 \caption{}
 \end{figure}

%
%

\begin{figure}
\figurenum{9}
\epsscale{0.7}
\plotone{f9.eps}
\vskip 1.0in
 \caption{}
 \end{figure}

%
%

\begin{figure}
\figurenum{10}
\epsscale{0.7}
\plotone{f10.eps}
\vskip 1.0in
 \caption{}
 \end{figure}

%
%

\begin{figure}
\figurenum{11}
\epsscale{1.0}
\plotone{f11.eps}
\vskip 0.5in
 \caption{}
 \end{figure}

%
%

\begin{figure}
\figurenum{12}
\epsscale{0.7}
\plotone{f12.eps}
\vskip 1.0in
 \caption{}
 \end{figure}

%
%

\begin{figure}
\figurenum{13}
\epsscale{0.9}
\vskip 0.5in \plotone{f13.eps}
\vskip 1.2in
 \caption{}
 \end{figure}

%
%

\begin{figure}
\figurenum{14}
\epsscale{1.0}
\plotone{f14.eps}
\vskip -0.2in
 \caption{}
 \end{figure}

\clearpage

%
%

\begin{deluxetable}{lcccc}
\scriptsize
\tablewidth{400pt}
\tablecaption{Program Stars.\label{tab:stars}}
\tablehead{
\colhead{Star} &
\colhead{R.A.\ (J2000)} &
\colhead{Dec.\ (J2000)} &
\colhead{$V$ (mag)} & 
\colhead{Exp.\ time (sec)} 
}
\startdata
OGLE-TR-3  & 17$^h$51$^m$48.$\!\!^s$95 &
$-$30$^\circ$13$^\prime$25.$\!\!^{\prime\prime}$1 & 16.6 & 1350 \\
OGLE-TR-10 & 17$^h$51$^m$28.$\!\!^s$25 &
$-$29$^\circ$52$^\prime$34.$\!\!^{\prime\prime}$9 & 15.8 & 1800, 2700\tablenotemark{a} \\
OGLE-TR-56 & 17$^h$56$^m$35.$\!\!^s$51 &
$-$29$^\circ$32$^\prime$21.$\!\!^{\prime\prime}$2 & 16.6 & 1350 \\
OGLE-TR-58 & 17$^h$55$^m$08.$\!\!^s$27 &
$-$29$^\circ$48$^\prime$51.$\!\!^{\prime\prime}$3 & 16.0 & 2700, 2$\times$1800\tablenotemark{a}\\
HD~179949  & 19$^h$15$^m$33.$\!\!^s$23 &
$-$24$^\circ$10$^\prime$45.$\!\!^{\prime\prime}$7 & \phn6.3 & 60, 60\tablenotemark{a} \\
HD~209458  & 22$^h$03$^m$10.$\!\!^s$80 &
$+$18$^\circ$53$^\prime$04.$\!\!^{\prime\prime}$0 & \phn7.7 & 60, 60\tablenotemark{a} \\
HIP   1078 & 00$^h$13$^m$24.$\!\!^s$64 &
$+$19$^\circ$04$^\prime$16.$\!\!^{\prime\prime}$8 & \phn9.9 & 300, 600\tablenotemark{a} \\
HIP   1334 & 00$^h$16$^m$43.$\!\!^s$37 &
$+$20$^\circ$51$^\prime$24.$\!\!^{\prime\prime}$1 & 10.0 & 300, 600\tablenotemark{a} \\
HIP   2102 & 00$^h$26$^m$40.$\!\!^s$48 &
$+$30$^\circ$11$^\prime$58.$\!\!^{\prime\prime}$0 & 10.1 & 300, 600\tablenotemark{a} \\
HIP   3232 & 00$^h$41$^m$07.$\!\!^s$90 &
$+$19$^\circ$15$^\prime$15.$\!\!^{\prime\prime}$9 & \phn9.9 & 300, 600\tablenotemark{a} \\
HIP 117177 & 23$^h$45$^m$24.$\!\!^s$64 &
$+$39$^\circ$07$^\prime$24.$\!\!^{\prime\prime}$2 & 10.2 & 300, 600\tablenotemark{a} \\
HIP 117846 & 23$^h$54$^m$04.$\!\!^s$83 &
$+$17$^\circ$33$^\prime$00.$\!\!^{\prime\prime}$2 & 10.1 & 300, 600\tablenotemark{a} \\
\enddata
\tablenotetext{a}{Template exposure for use in the iodine reductions.}
\end{deluxetable}

\clearpage

%
%

\begin{deluxetable}{lccc}
\scriptsize
\tablewidth{270pt}
\tablecaption{Thorium-Argon radial velocities in the barycentric frame for
the OGLE targets.\label{tab:thar_rvs}}
\tablehead{
\colhead{HJD} & 
\colhead{Velocity ($\kms$)} & 
\colhead{Error ($\kms$)} }
\startdata
\cutinhead{OGLE-TR-3}
2452481.87318 & $-$23.03 & 0.35 \\
2452483.87185 & $-$23.40 & 0.07 \\
\cutinhead{OGLE-TR-56}                            
2452480.92390 & $-$49.26 & 0.20 \\
2452481.90961 & $-$49.35 & 0.07 \\
2452483.90680 & $-$49.64 & 0.07 \\
\enddata
\end{deluxetable}

\clearpage

%
%

\begin{deluxetable}{lccc}
\scriptsize
 \tablewidth{270pt} 
\tablecaption{Iodine radial velocities for the OGLE targets.\tablenotemark{a}\label{tab:iodine_rvs}}
 \tablehead{
\colhead{HJD} & 
\colhead{Velocity ($\kms$)} & 
\colhead{Error ($\kms$)} }
\startdata
\cutinhead{OGLE-TR-10}
2452481.79455 & $+$0.09 & 0.06 \\
2452483.75773 & $-$0.09 & 0.05 \\
\cutinhead{OGLE-TR-58}                            
2452481.83607 & $-$0.18 & 0.16 \\
2452483.79181 & $+$0.18 & 0.06 \\
\enddata
\tablenotetext{a}{These velocities are referred to the average for each star, 
since the iodine technique yields only relative velocities.}
\end{deluxetable}

\end{document}